\documentclass[preprint,aps,floatfix,nofootinbib,a4paper, superscriptaddress]{revtex4-1}

\usepackage[utf8]{inputenc}

\usepackage{amsmath, amssymb, amsfonts, amsthm, latexsym, epsfig, mathrsfs, xcolor, bbm, slashed, xypic}
\usepackage[colorlinks, allcolors=blue!70!black, linktocpage]{hyperref}
\usepackage{enumitem}

\let\OLDthebibliography\thebibliography
\renewcommand\thebibliography[1]{
  \OLDthebibliography{#1}
	\setlength{\baselineskip}{3.236ex plus 0.3ex} 
  \setlength{\itemsep}{1.618ex plus 0.1ex} 
}

\renewcommand\thesection{\arabic{section}}
\renewcommand\thesubsection{\arabic{subsection}}

\makeatletter
\def\p@subsection{\thesection.}
\def\p@subsubsection{\thesection.\thesubsection.}
\makeatother 

\numberwithin{equation}{section}

\theoremstyle{definition}
\newtheorem{thm}{Theorem}

\newtheorem{prop}{Proposition}[section]

\newtheorem{remark}{Remark}

\theoremstyle{definition}

\newcommand{\be}{\begin{equation}}
\newcommand{\ee}{\end{equation}}

\newcommand{\lb}{\left}
\newcommand{\rb}{\right}

\newcommand{\mc}{\mathcal}

\newcommand{\ms}{\mathscr}
\newcommand{\mf}{\mathfrak}

\newcommand{\eqsp}{\hspace{10pt};\hspace{10pt}} 

\newcommand{\norm}[1]{\lb\Vert\, #1 \,\rb\Vert}		
\newcommand{\inp}[1]{\lb\langle #1 \rb\rangle}		

\newcommand{\sqh}{\sqrt{h}}  

\newcommand{\Lie}{\pounds} 
\newcommand{\Lap}{D^2} 

\newcommand{\defn}{\mathrel{\mathop:}=} 

\newcommand{\df}[1]{\boldsymbol{#1}} 

\begin{document}

\pagestyle{myheadings} 


\title{A Variational Principle for the Axisymmetric Stability of Rotating Relativistic Stars}

\author{Kartik Prabhu}\email{kartikp@uchicago.edu}
\affiliation{Enrico Fermi Institute and Department of Physics, The University of Chicago, Chicago, IL 60637, USA}
\author{Joshua S. Schiffrin}\email{schiffrin@damtp.cam.ac.uk}
\affiliation{DAMTP, University of Cambridge, Centre for Mathematical Sciences, Wilberforce Road, Cambridge CB3 0WA, UK}
\author{Robert M. Wald}\email{rmwa@uchicago.edu}
\affiliation{Enrico Fermi Institute and Department of Physics, The University of Chicago, Chicago, IL 60637, USA}

\begin{abstract}
	It is well known that all rotating perfect fluid stars in general relativity are unstable to certain non-axisymmetric perturbations via the Chandrasekhar-Friedman-Schutz (CFS) instability. However, the mechanism of the CFS instability requires, in an essential way, the loss of angular momentum by gravitational radiation and, in many instances, it acts on too long a timescale to be physically/astrophysically relevant. It is therefore of interest to examine the stability of rotating, relativistic stars to axisymmetric perturbations, where the CFS instability does not occur. In this paper, we provide a Rayleigh-Ritz type variational principle for testing the stability of perfect fluid stars to axisymmetric perturbations, which generalizes to axisymmetric perturbations of rotating stars a variational principle given by Chandrasekhar for spherical perturbations of static, spherical stars. Our variational principle provides a lower bound to the rate of exponential growth in the case of instability. The derivation closely parallels the derivation of a recently obtained variational principle for analyzing the axisymmetric stability of black holes.

\end{abstract}

\maketitle
\tableofcontents


\section{Introduction}\label{sec:intro}

Based on the work of Chandrasekhar \cite{Chandrasekhar}, and Friedman and Schutz \cite{FriedmanSchutz}, Friedman \cite{Friedman} developed a canonical energy method for studying the linear stability of perfect fluid stars in general relativity.\footnote{\footnotesize In this paper we will be concerned with dynamic stability of stars; the criterion for thermodynamic stability of relativistic stars to axisymmetric perturbations is given in \cite{GSW} (see also \cite{LH}).} The method consists of using a Lagrangian formulation of the Einstein-perfect fluid system to define a quantity, $\mathcal E(\delta X)$, known as the canonical energy, that is given by an integral over a Cauchy surface $\Sigma$ of an expression that is bilinear in the perturbed initial data, $\delta X$. It can then be shown that $\mathcal E(\delta X)$ is gauge invariant, conserved (i.e., independent of the choice of $\Sigma$), and has positive flux at null infinity. Furthermore, when restricted to a certain subspace \(\ms V_c\) of perturbations that have vanishing Lagrangian change in circulation,\footnote{\footnotesize Vanishing Lagrangian change in circulation is a necessary condition for perturbations to be in the subspace \(\ms V_c\), but it is not sufficient; see \cite{GSW} for the full discussion.\medskip} $\mathcal E(\delta X)$ is degenerate on, and only on, perturbations to other stationary solutions. It follows that if $\mathcal E(\delta X)$ is always positive on $\ms V_c$, then it provides a conserved norm that excludes the possibility of mode instability. On the other hand if there exists a perturbation $\delta X \in \ms V_c$ for which $\mathcal E(\delta X) < 0$, then this perturbation must be unstable in the sense that it cannot settle down to a stationary solution at late times, since the positive flux property implies that the canonical energy of this limiting stationary solution be must strictly negative, in contradiction with the fact that the canonical energy vanishes for stationary perturbations. Thus, positivity of $\mathcal E$ is a general criterion for stability. Friedman \cite{Friedman} was further able to show that for any rotating perfect fluid star in general relativity, one always can find perturbations $\delta X$ with angular dependence $e^{i m \phi}$ for large enough $m$ such that $\mathcal E(\delta X) < 0$. Thus, all rotating stars are unstable (the CFS instability). 

The canonical energy method for showing existence of an instability has a great advantage over the straighforward approach of finding growing solutions to the linearized field equations, since one need not solve the full set of linearized equations; rather, one only needs to find a solution, $\delta X$, of the linearized initial-value constraint equations that has negative canonical energy. However, the canonical energy method directly shows instability only in the weak sense of the previous paragraph --- the impossibility of settling down to a stationary end-state --- rather than proving the existence of an exponentially growing mode. Furthermore, if a perturbation $\delta X$ is found with $\mathcal E(\delta X) < 0$, there is no information directly available from the canonical energy method on the growth rate of the instability. Indeed, for stars that are not highly relativistic and rapidly rotating, the growth timescale for the CFS instability is expected to be longer than astrophysically relevant timescales. However, there is no known way of determining the growth timescale of the CFS instability from the canonical energy method.

Several years ago, the canonical energy method was extended to the case of vacuum black holes in arbitrary dimensions \cite{HW-stab}. In this case, it was necessary to restrict consideration to axisymmetric perturbations because there are now two boundaries through which canonical energy can pass: null infinity and the black hole horizon. As in the fluid star case, the net flux of canonical energy through null infinity is positive if canonical energy is defined with respect to the Killing field of the background spacetime that is timelike at infinity. However, the net flux of canonical energy through the horizon is similarly positive only when it is defined relative to the horizon Killing field. If the black hole is static, then these Killing fields coincide, and one can make the same type of arguments as above. However, for a rotating black hole, these two notions of canonical energy agree only for axisymmetric perturbations, so one may make the above stability arguments only for the case of axisymmetric perturbations.\footnote{\footnotesize For the case of asymptotically-AdS black holes, there is again only one boundary through which canonical energy can pass---namely, the black hole horizon---so one need not restrict to axisymmetric perturbations in this case, and one can prove that any asymptotically-AdS black holes with an ergoregion must be unstable \cite{GHIW}.\medskip}

Recently, it was shown that for arbitrary perturbations of static black holes and for axisymmetric perturbations of rotating black holes, the canonical energy approach can be extended so as to obtain information on the rate of exponential growth of instabilities. The key idea in this extension is to break up a perturbation into its odd and even parts under the $t$ or $t$-$\phi$ reflection isometry of the background solution \cite{Pap, Carter-killing, SW-tphi}. The canonical energy will correspondingly break up into a sum of two pieces, which we refer to, respectively, as the ``kinetic energy'' and ``potential energy'' of the perturbation. It was shown in \cite{PW} that the kinetic energy is always positive. Therefore, an instability can occur only if the potential energy can be made negative. The main result then proven in \cite{PW} is that if the potential energy can be made negative for a perturbation that can be expressed as the time derivative of another perturbation, then that perturbation must grow exponentially with time. Furthermore, a Rayleigh-Ritz type of variational principle can be given, which provides a rigorous lower bound on the rate of exponential growth.\footnote{\footnotesize In the astrophysically relevant case of $4$-spacetime dimensions, the only black hole solutions are the Kerr family of metrics, which are believed to be stable \cite{Whiting-mode-stab}, so there is presumably no need for a method to bound exponential growth rates. However, the variational principle of \cite{PW} applies to black holes and black branes in arbitrary dimensions, where instabilities do occur.}

The purpose of this paper is to extend the variational principle results of \cite{PW} to the case of perfect fluid stars in general relativity that are either static, or stationary and axisymmetric with circular flow. Since no horizon is present, the canonical energy method by itself does not require us to restrict consideration to axisymmetric perturbations. However, for non-axisymmetric perturbations of stationary-axisymmetric rotating stars, the CFS instability implies that the kinetic energy cannot be positive definite.\footnote{\footnotesize For a perturbation with ``angular quantum number'' $m \neq 0$, a rotation in $\phi$ by $\pi/2m$ will take a $t$-$\phi$ odd perturbation to a $t$-$\phi$ even perturbation. It follows that for non-axisymmetric perturbations of a rotating star, the kinetic energy cannot be positive definite unless the full canonical energy is positive definite.\medskip} Since positivity of kinetic energy is the key property needed to establish exponential growth and obtain a Rayleigh-Ritz type of variational principle, our results in this regard will apply only to axisymmetric perturbations of stationary-axisymmetric stars; for static stars we will not require this restriction. Our analysis will closely follow \cite{PW}, but with significant simplifications from the absence of a black hole horizon and significant complications from the nature of the Lagrangian formulation of perfect fluids.

	The remainder of this paper is organized as follows. In \autoref{sec:bg} we review the Lagrangian formulation of perfect fluids and describe the background spacetime of interest. In \autoref{sec:lin-pert} we review the symplectic structure and canonical energy of linearized perturbations of the Einstein-fluid star and define the space of perturbations that we consider. In \autoref{sec:positive-KE}, we split the canonical energy into ``kinetic" and ``potential" parts and prove that the kinetic energy is positive definite. In \autoref{sec:exp-growth}, we use the positivity of kinetic energy to show that negative potential energy implies the existence of an exponentially growing perturbation, and we derive our variational formula for the growth rate. In \autoref{sec:algorithm} we provide an explicit algorithm to compute the variational formula and show that, for spherically-symmetric perturbations of static spherically-symmetric stars with a ``barotropic" fluid equation of state, it reduces to that of \cite{Chandra1, Chandra, SW}.

We will use an abstract index notation for tensor fields. Greek letters \(\mu,\nu,\ldots\) denote tensors on spacetime \(M\) (e.g. \(u^\mu\) is the \(4\)-velocity vector) while, Latin letters \(a,b,\ldots\) denote tensors on a spacelike hypersurface \(\Sigma\) (e.g. \(u^a\) is the projection of $u^\mu$ into $\Sigma$). Differential forms will be denoted by a bold-face when using an index-free notation (e.g. \(\df N\) is the particle current \(3\)-form).

\section{Background Spacetime}\label{sec:bg}

A Lagrangian formulation of the Einstein-perfect fluid system was described in \cite{Friedman,FS-book, GSW}. In this formulation, one introduces a fiducial manifold \(M'\) that is diffeomorphic to the spacetime  \(M\). Further, one chooses on \(M'\) a fixed scalar field \(s'\) and a fixed \(3\)-form \(\df{N}'\) such that
\be\label{dNeqn}
	d(\df{N}') = d(s'\df{N}') = 0 \ .
\ee
The dynamical fields are given by the pair 
\be
\Psi \defn (g_{\mu\nu}, \chi) \, ,
\ee
where $g_{\mu\nu}$ is a spacetime metric on $M$ and \(\chi: M' \to M\) is a diffeomorphism. 

The physical variables of the fluid are then obtained from these dynamical variables as follows: The pushforwards \(s = \chi_* s'\) and \(\df{N} = \chi_* \df{N}'\) are, respectively, the \emph{entropy per particle} and the \emph{particle current} \(3\)-form. The \emph{particle number density}, \(n\), and the \emph{fluid 4-velocity}, \(u^\mu\), are then given by the relations\footnote{\footnotesize It is assumed/required here that the diffeomorphism $\chi$ is such that $N_{\mu\nu\lambda}N^{\mu\nu\lambda} \geq 0$.\medskip}
\be
	n = \left(\frac{1}{3!}N_{\mu\nu\lambda}N^{\mu\nu\lambda}\right)^{1/2} \eqsp N_{\mu\nu\lambda} = n u^\rho \varepsilon_{\rho\mu\nu\lambda} \, .
\ee
It follows that $n\geq0$ and that \(u^\mu\) is a unit future-directed time-like vector field. The \emph{energy density}, $\rho$, is assumed to be given in terms of $n$ and $s$ by specifying an equation of state, 
\be
\rho = \rho(n, s) \, .
\ee
The \emph{pressure}, $p$, is then given by the thermodynamic relation 
\be\label{pressuredefn}
p= n \frac{\partial \rho}{\partial n}-\rho \, .
\ee 
We assume that the equation of state is chosen so that 
\be\label{fluidconditions}
\rho \geq 0 \eqsp p\geq 0 \eqsp  0\leq c_s^2 \leq 1\, 
\ee 
where
\be\label{cdefn}
c_s^2 \defn \frac{\partial p/\partial n}{\partial \rho/ \partial n} \, .
\ee
Under these conditions, the Einstein-perfect fluid equations are well-posed (see, e.g., \cite{CBYork}).  

The Lagrangian 4-form for the Einstein-perfect fluid system is
\be\begin{split}
	\df L & = \df L_{\rm GR} + \df L_{\rm fluid} \\
		& = \lb[\frac{1}{16\pi}R - \rho(n, s) \rb] \df \varepsilon \, .
\end{split}\ee
Varying with respect to the dynamical fields \(\Psi=(g_{\mu\nu}, \chi)\) gives the Einstein-fluid equations of motion:\footnote{\footnotesize Of course, the second equation follows from the first. Outside of the Lagrangian formulation there is an additional field equation, namely conservation of particle number, $\nabla_\mu (n u^\mu) = 0$, but in the Lagrangian formulation this follows automatically from \autoref{dNeqn}. Similarly conservation of entropy, $\nabla_\mu (sn u^\mu)$, follows from \autoref{dNeqn}. The $u^\nu$-component of \autoref{eq:consT} follows automatically from conservation of particle number and entropy, and the remaining components give the Euler equation \autoref{eq:euler-eqn}.\medskip}
\begin{subequations}\label{eq:Ein-fluid}\begin{align}
	- \tfrac{1}{16\pi}G^{\mu\nu} + \tfrac{1}{2} T^{\mu\nu} & = 0 \label{eq:Ein}\\
	- \nabla^\mu T_{\mu\nu} & = 0 \,  \label{eq:consT}
\end{align}\end{subequations}
where
\be\label{eq:T-fluid}
	T^{\mu\nu} = (\rho + p) u^\mu u^\nu + p g^{\mu\nu} \, .
\ee
is the perfect fluid stress-energy tensor.
\\

We consider globally hyperbolic, asymptotically flat solutions of the Einstein-perfect fluid equations in \((3+1)\)-dimensions that represent a ``star'' in \emph{dynamic equilibrium}, i.e., solutions for which $\df N$ has compact spatial support that are either static or stationary-axisymmetric. In the static case, the spacetime possesses a $t$-reflection symmetry, by definition. In the stationary-axisymmetric case we only consider solutions having \emph{circular flow}, meaning that the fluid velocity lies in the plane spanned by the Killing fields, i.e., 
\be \label{circflow}
u^\mu = \frac{t^\mu + \Omega \phi^\mu}{V}
\ee
where \(V\) is a normalization factor so that \(u^\mu\) is unit-time-like. It then follows that the spacetime possesses a ($t$-$\phi$)-reflection symmetry \cite{Pap, Carter-killing}.\footnote{\footnotesize In higher dimensions, the arguments of \cite{Pap, Carter-killing} cannot be applied, but the existence of  a ($t$-$\phi$)-reflection symmetry can be shown by the arguments given in \cite{SW-tphi, Sch}.\medskip} The $t$ or $t$-$\phi$ reflection symmetries play a key role for our results, as they will allow us to define a preferred decomposition of the canonical energy into kinetic and potential parts.  

Let \(\Sigma\) be a Cauchy surface for the spacetime. Below, we will choose $\Sigma$ to be invariant under the reflection isometry but we need not make this choice now, and it will be convenient not to do so until later so that the variations of our equations will hold for general perturbations, where the reflection isometry need not be present. Let  \( \Sigma_t \) denote the foliation obtained by applying time translations to \(\Sigma\). Let $\nu^\mu$ denote the future-directed unit normal to \(\Sigma\). We decompose the time-translation Killing vector field, $t^\mu$, into its normal and tangential parts relative to $\Sigma$, referred to as the \emph{lapse}, \(N = - \nu_\mu t^\mu\), and \emph{shift}, \(N^a\), on \(\Sigma\).

Let \( h_{ab} \) denote the induced metric on $\Sigma$ and let \(K_{ab}\) denote the extrinsic curvature of $\Sigma$. The initial data for the gravitational variables on \(\Sigma\) is given by \(\lb( \pi^{ab}, h_{ab}\rb) \) where \(\pi^{ab} \defn \sqh(K^{ab}-K~h^{ab}) \) is the canonical momentum-density conjugate to \(h_{ab}\). In order to correspond to a solution of Einstein's equation \autoref{eq:Ein}, the initial data must satisfy the constraint equations \( {\mf C}^\mu= 0\), with
\begin{subequations}\label{eq:constraints}
	\begin{align}
		\mf C  \defn \nu_\mu \mf C^\mu & = \tfrac{1}{16\pi} \lb[ - R + \tfrac{1}{h} \lb( \pi^{ab}\pi_{ab} - \tfrac{1}{2}({\pi_a}^a)^2\rb) \rb] + E \label{eq:constraint-H}	\\
		\mf C^a \defn {h_\mu}^a \mf C^\mu & = \tfrac{1}{16\pi} \lb[ -2D_b\lb( \frac{\pi^{ab}}{\sqh} \rb) \rb] - \frac{J^a}{\sqh}  \label{eq:constraint-diff}
	\end{align}
\end{subequations}
where \( D_a \) denotes the covariant derivative on \(\Sigma\) compatible with \( h_{ab} \), and \( R_{ab} \) is the Ricci curvature of \(h_{ab}\). The matter contributions are defined as
\begin{subequations}\label{eq:rho-J}\begin{align}
	E &\defn \nu_\mu \nu_\nu T^{\mu\nu} \label{eq:rho}\\
	J^a &\defn - \sqh \nu_\mu {h_\nu}^{a}T^{\mu\nu}
\end{align}\end{subequations}
where ${h_\mu}^{a}$ is the projection tensor into $\Sigma$.

The ADM time evolution equations for the gravitational initial data are (see Sec.VI.6 \cite{CB-book}):\footnote{\footnotesize Note that we have set Newton's constant \(G_N = 1\), while \cite{CB-book} uses the convention that \(8\pi G_N = 1\). Also, following \cite{Wald-book} we define the extrinsic curvature by \(K_{ab} \defn \tfrac{1}{2}\Lie_\nu h_{ab}\) which differs by a sign from \cite{CB-book}.\medskip\medskip}
\begin{subequations}\label{eq:evol-bg}
	\begin{align}
		\begin{split}		
		\tfrac{1}{\sqh}\dot\pi^{ab}  & =  -N \lb( R^{ab} - \tfrac{1}{2} R h^{ab}\rb) + \tfrac{N}{2h}~h^{ab} \lb( \pi_{cd}\pi^{cd} - \tfrac{1}{2}({\pi_c}^c)^2 \rb) - \tfrac{2N}{h}\lb( \pi^{ac} {\pi_c}^b - \tfrac{1}{2} {\pi_c}^c\pi^{ab} \rb) \\
		&~\quad +  D^aD^bN - h^{ab}\Lap N  + D_c \lb( \tfrac{1}{\sqh}\pi^{ab} N^c \rb)   - \tfrac{2}{\sqh}~ \pi^{c(a}D_c N^{b)} + 8 \pi N T^{ab}
		\end{split} \label{eq:evol-bg-pi}\\ 
		\dot h_{ab}  & = \frac{2N}{\sqh}\lb( \pi_{ab} - \tfrac{1}{2}{\pi_c}^c h_{ab} \rb) + 2 D_{(a}N_{b)} \label{eq:evol-bg-h}
	\end{align}
\end{subequations}
where the overdot denotes $\Lie_t$ and 
\be
\Lap \defn D^aD_a 
\ee
is the Laplacian on \(\Sigma\), and \(T^{ab} \defn {h_\mu}^{a}{h_\nu}^{b}T^{\mu\nu}\). Since we are considering stationary background spacetimes, the left side of \autoref{eq:evol-bg} vanishes in the background.

 In terms of the fluid variables, the matter contributions to the constraints and ADM equations are
\begin{subequations}\label{eq:fluid-matter}\begin{align}
	E & = \rho + (\rho + p) u^2 \label{eq:fluid-rho} \\
	 J^a & = \sqh(\rho + p ) \sqrt{1+u^2}~ u^a  \label{eq:fluid-J}\\
	 T^{ab} & = (\rho + p)u^a u^b + p h^{ab} \label{eq:fluid-L}
\end{align}\end{subequations}
where \(u^a \defn {h_\mu}^a u^\mu \) and \(u^2 \defn  h_{ab}u^a u^b \) so that \(\nu_\mu u^\mu = - \sqrt{1+u^2}\).

	The projection to $\Sigma$ of the Euler equation, $h_\nu{}^a \left( {\delta^\nu}_\sigma + u^\nu u_\sigma \right) \nabla_\rho T^{\sigma\rho}=0$, can be written as
\be\begin{aligned}\label{eq:euler-eqn}
	\dot u^a &= \Lie_N u^a - \sqrt{1+u^2}\, D^a N -\tfrac{2N}{\sqrt{h}} \pi^{ab} u_b +\tfrac{N}{\sqrt{h}} {\pi^b}_b u^a- \tfrac{N}{\sqrt{1+u^2}} u^b D_b u^a \\
	&\qquad - \tfrac{1}{(\rho+p)}\left[  \tfrac{N}{\sqrt{1+u^2}}\left(h^{ab}+ u^a u^b \right)D_b p +u^a \left( \dot p - N^b D_b p \right)  \right].
\end{aligned}\ee\\

We can significantly further simplify the right side of \autoref{eq:evol-bg} by choosing $\Sigma$ to be invariant (i.e., mapped into itself) under the $t$-reflection isometry (in the static case) or the ($t$-$\phi$)-reflection isometry (in the stationary-axisymmetric case). We first discuss the stationary-axisymmetric case, and then make the additional simplifications that occur in the static case. In the stationary-axisymmetric case, $\Sigma$ is obtained by taking the orbits under the action of the axial Killing field $\phi^\mu$ of the $2$-dimensional surfaces orthogonal to $t^\mu$ and $\phi^\mu$. It follows immediately that $\phi^\mu$ is tangent to $\Sigma$, so we may denote it as $\phi^a$. The restriction to $\Sigma$ of the ($t$-$\phi$)-reflection isometry, $i$, then maps the $2$-surfaces orthogonal to $\phi^a$ to themselves and satisfies 
\be
i^* \phi^a = -\phi^a \eqsp i^* h_{ab} = h_{ab}  \eqsp i^* \pi^{ab} = -\pi^{ab}.
\ee
Since $\pi^{ab}$ is odd under $i^*$, it follows that $\pi^{ab}$ takes the form
\be
\pi^{ab} = 2\sqh \pi^{(a}\phi^{b)}
\ee
with $\pi_a\phi^a = 0$. Since $t^\mu$ is odd under the ($t$-$\phi$)-reflection isometry, it also follows that 
\be
t^\mu = N \nu^\mu + \bar N \phi^\mu,
\ee
i.e., the shift vector takes the form $N^a = \bar N \phi^a$. Finally, since $\phi^a$ is $2$-surface orthogonal Killing field on $\Sigma$, we have
\be\label{eq:Dphi}
		D_a\phi_b = - \Phi^{-1} \phi_{[a}D_{b]}\Phi \, 
	\ee
where
\be
\Phi \defn h_{ab} \phi^a\phi^b.
\ee

Since the fluid flow is circular it follows that
\be\label{eq:circular-flow}
	u^a = U \phi^a \eqsp u^2 = \Phi U^2
\ee
and from \autoref{eq:fluid-matter} that
\begin{subequations}\label{eq:matter-tphi}\begin{align}
	J^a & = \bar J\phi^a \\
	T^{ab} & = \hat T^{ab} + \bar T \phi^a\phi^b
\end{align}\end{subequations}
where \(\hat T^{ab}\phi_a = 0\). In other words, $J^a$ is ``axial'' (odd under $\phi$-reflection) while $T^{ab}$ is ``polar'' (even). 
\\

With the above choice of $\Sigma$, the constraint equations \autoref{eq:constraints} become:
\begin{subequations}\label{eq:constraints-ax}\begin{align}
	R & = 2\Phi \pi_a \pi^a + 16 \pi E \label{eq:constraints-H-ax} \\
	D_a\lb(\Phi \pi^a\rb) &=- 8 \pi \Phi \bar J. \label{eq:constraints-diff-ax}
\end{align}\end{subequations}
In addition, the ADM evolution equations \autoref{eq:evol-bg} can be simplified. \autoref{eq:evol-bg-h} becomes
\be\label{eq:shift-id}
D_a \bar N  = - 2 N\pi_a.
\ee
Using this we have
\be\begin{split}
	D_c \lb( \tfrac{1}{\sqh}\pi^{ab} N^c \rb)  - \tfrac{2}{\sqh}~ \pi^{c(a}D_c N^{b)}  & = 4 N \pi^{c} \pi_c\phi^a \phi^b.
\end{split}\ee
Taking the trace of \autoref{eq:evol-bg-pi} and using the using \autoref{eq:constraints-ax} we have
\be\label{eq:Lap-N-id}
	\Lap N = 2N (\Phi \pi_a \pi^a) + 4\pi N \lb( E + {T_a}^a\rb) .
\ee
Hence, \autoref{eq:evol-bg-pi} simplifies to
\be\label{eq:Ric-id}\begin{split}
NR_{ab} & = D_aD_b N -2 N\lb( \Phi \pi_a\pi_{b} - \pi_c \pi^{c} \phi_{a}\phi_{b} \rb) + 8 \pi N V_{ab}
\end{split}\ee
where
\be\label{eq:V-fluid-defn}
	V_{ab} \defn T_{ab} + \frac{1}{2}h_{ab}(E-{T_c}^c).
\ee

Similarly, the Euler equation \autoref{eq:euler-eqn} reduces to
\be\label{eq:dp-station}
	\frac{N}{\rho+p} D_a p = - (1+\Phi U^2)~ D_a N - 2N U \sqrt{1+\Phi U^2}~ \pi_a + \frac{N}{2} U^2 D_a \Phi.
\ee\\

These relations simplify considerably in the static case, where $\pi^{ab} = 0$ and $N^a = u^a = 0$. The ADM evolution equations then reduce to 
\be\label{eq:Ric-id-static}
NR_{ab} = D_aD_bN + 8\pi N V_{ab} \,,
\ee
the constraint equations reduce to
\be
R = 16 \pi E\, ,
\ee
while the Euler equation becomes
\be\label{eq:dp-static}
	\frac{N}{\rho+p} D_a p = - D_a N.
\ee

\section{Linear Perturbations: Symplectic Structure and Canonical Energy}\label{sec:lin-pert}

Consider a one-parameter family of Einstein-fluid spacetimes given by the dynamical fields \(\Psi(\lambda)=(g_{\mu\nu}(\lambda), \chi(\lambda))\), which is smooth in both \(\lambda\) and on \(M\), with \(\Psi(0)\) corresponding to a static star or a stationary-axisymmetric star with circular flow. A linearized perturbation is then described by the perturbed metric $\delta g_{\mu \nu} \defn
(d g_{\mu \nu}/d \lambda) |_{\lambda=0}$ together with the vector field \(\xi^\mu\)---called the \emph{Lagrangian displacement}--- that is the infinitesimal generator of the one-parameter family of diffeomorphisms \(\chi(\lambda) \circ \chi(0)^{-1}: M \to M\) (see, e.g., \cite{FS-book, GSW}). Perturbed physical fluid quantities are then obtained using
\be\label{deltaNs}
\delta \df N = -\Lie_\xi \df N \eqsp \delta s = -\Lie_\xi s.
\ee 
It is useful to define the \emph{Lagrangian perturbation} of a quantity $Q$ as
\be\label{Deltadefn}
\Delta Q \defn \delta Q + \Lie_\xi Q \, ,
\ee
so that $\Delta Q$ corresponds to the perturbation of $Q$ in a gauge where $\xi^\mu = 0$.
From \autoref{deltaNs}, we then have $\Delta \boldsymbol N = \Delta s = 0$. One finds the Lagrangian perturbation of $n$ and $u^\mu$ (see \cite{FS-book, GSW}) to be given by
\begin{subequations}
\begin{align}
\Delta n &= -\frac{1}{2} n (g^{\mu\nu} - u^\mu u^\nu) \Delta g_{\mu\nu} \\
\Delta u^\mu &=\frac{1}{2}u^\mu u^\nu u^\lambda \Delta g_{\nu\lambda}\label{deltau}
\end{align}
\end{subequations}
with 
\be
\Delta g_{\mu\nu} = \delta g_{\mu\nu} + 2 \nabla_{(\mu}\xi_{\nu)}.
\ee
It will be useful to write, using \autoref{Deltadefn} and $\Delta s=0$,
\be\begin{aligned}\label{deltarhoandp}
\delta \rho &= \frac{\rho+p}{n} \Delta n -\Lie_\xi \rho\\
\delta p &= c_s^2 \frac{\rho+p}{n} \Delta n -\Lie_\xi p,\\
\end{aligned}\ee
where we have used \autoref{pressuredefn} and \autoref{cdefn} to write
\be\label{partials}
\frac{\partial \rho}{\partial n}  = \frac{\rho+p}{n} \eqsp 
\frac{\partial p}{\partial n}  = c_s^2 \frac{\rho+p}{n}.
\ee

The perturbed gravitational initial data is given by
\be
\sqh p^{ab} \defn \delta\pi^{ab} = \lb.\frac{d}{d\lambda}\pi^{ab}(\lambda)\rb\vert_{\lambda=0} \eqsp q_{ab} \defn \delta h_{ab} = \lb.\frac{d}{d\lambda}h_{ab}(\lambda)\rb\vert_{\lambda=0}.
\ee
On the Cauchy surface \(\Sigma\), the symplectic form \(W_\Sigma\) for two perturbations \(\delta \Psi\), \(\widetilde{\delta \Psi}\) of the Einstein-fluid system is given by (see \cite{HW-stab, GSW})
\be\label{eq:symp-form}\begin{split}
	W_\Sigma[\delta \Psi,\widetilde{\delta \Psi}] & = W_{\rm GR}[\delta \Psi,\widetilde{\delta \Psi}] + W_{\rm fluid}[\delta \Psi,\widetilde{\delta \Psi}] \\
		&= \frac{1}{16\pi}\int_\Sigma \df\varepsilon^{(3)}\lb( p_{ab} \tilde q^{ab} - q^{ab} \tilde p_{ab} \rb) + \int_\Sigma \left( \tilde \xi^\mu\delta P_{\mu\nu\lambda\rho}  -  \xi^\mu \widetilde{\delta P}_{\mu\nu\lambda\rho} - [\xi,\tilde\xi]^\mu P_{\mu\nu\lambda\rho}\right)
\end{split}\ee
where \(\df\varepsilon^{(3)}\) is the background volume form on \(\Sigma\) and
\be\label{eq:P-defn}
	P_{\mu\nu\lambda\rho} \defn (\rho+p) ({\delta_\mu}^\sigma + u_\mu u^\sigma)\varepsilon_{\sigma\nu\lambda\rho}\,.
\ee
Henceforth we will work on a fixed choice of Cauchy surface, and so we drop \(\Sigma\) from the symplectic form and the integrals.

A perturbation of the form \((\delta g_{\mu\nu}=0, \xi^\mu=\eta^\mu)\) is called \emph{trivial} if it does not change the physical fluid variables, i.e., if $\eta^\mu$ is such that
\be\label{eq:trivial}
	\delta s = - \Lie_\eta s = 0  \eqsp \delta \df N = -\Lie_\eta \df N = 0.
\ee
We refer to such an $\eta^\mu$ as a \emph{trivial displacement}. Any trivial displacement takes the general form (see \cite{GSW})
\be\label{eq:triv-gen}
	\eta^\mu = f u^\mu + \tfrac{1}{n^2}N^{\mu\nu\lambda}\nabla_\nu Z_\lambda
\ee
where \(f\) is any function on spacetime and the one-form \(\df Z \equiv Z_\mu\) satisfies
\be\label{eq:triv-Z-form}
	u^\mu Z_\mu = 0 \eqsp \Lie_u \df Z = 0 \eqsp ds \wedge d \df Z = 0.
\ee
	A trivial displacement of the form \(\eta^\mu = fu^\mu\), where \(f\) is any function of spacetime, is called a \emph{flowline trivial}. As explained in Sec.~4.3 of \cite{GSW}, any trivial perturbation of the form \((\delta g_{\mu\nu}=0, f u^\mu)\), is a degeneracy of the symplectic form \autoref{eq:symp-form} (even when the linearized constraints are not assumed to hold for the perturbations with which the symplectic product is being taken). Thus, we can always add a flowline trivial to any perturbation to make $\xi^\mu$ (and, in fact, any number of time derivatives of $\xi^\mu$) tangent to $\Sigma$ without affecting the symplectic form or any physical quantities. We take the perturbed initial data for the Einstein-fluid system to be
\be\label{eq:init-data-V}
\delta X \defn \left( p_{ab}, q_{ab}, v^a, \xi^a\right)
\ee 
where
\be
\xi^a \defn {h_\mu}^a \xi^\mu
\ee
and 
\be
v^a \defn \delta u^a = \delta\left({h_\mu}^a u^\mu \right)
\ee
are vector fields on $\Sigma$. Note, however, that $\xi^a$ and $v^a$ are not canonically conjugate with respect to the symplectic form \autoref{eq:symp-form}; see \cite{GSW} for a full discussion of the phase space.

We are interested in the space \(\ms P\) consisting of smooth and asymptotically flat perturbed initial data \(\delta X\). To correspond to solutions of the linearized Einstein-fluid equations the perturbed initial data must satisfy the linearized constraints \(\mf c^\mu \defn \delta \mf C^\mu = 0\). Linearizing \autoref{eq:constraints} we can write these as
\begin{subequations}\label{eq:lin-constraints}\begin{align}
	\begin{split}
	\mf c & = \tfrac{1}{16\pi}\lb[ \tfrac{2}{\sqh}\lb( \pi^{ab} - \tfrac{1}{2}{\pi_c}^c~ h^{ab} \rb)p_{ab} + \tfrac{2}{h}\lb( \pi^{ac} {\pi_c}^b - \tfrac{1}{2} {\pi_c}^c\pi^{ab} \rb) q_{ab} \rb. \\
		&~\quad \lb. -\tfrac{1}{h}\lb( \pi^{ab}\pi_{ab} - \tfrac{1}{2}({\pi_a}^a)^2 \rb){q_c}^c -  D^aD^bq_{ab} + \Lap {q_c}^c + R^{ab}q_{ab} \rb] + \delta E 
	\end{split}\label{hamconstraint} \\[15pt]
	\mf c^a & = \tfrac{1}{16\pi}\lb [-2D_bp^{ab} - \tfrac{1}{\sqh}\pi_{bc} \lb ( D^c q^{ba} + D^b q^{ca} - D^a q^{bc} \rb) \rb] - \tfrac{1}{\sqh} \delta J^a \label{momconstraint}
\end{align}\end{subequations}
where \( \delta J^a\) and \(\delta E\) can be obtained from \autoref{eq:fluid-matter}.

The canonical energy \(\ms E\) is a quadratic form on  \(\ms P\) defined in terms of the symplectic form by
\be\label{eq:canonical-energy}
	\ms E(\delta X, \widetilde{\delta X}) \defn W[\delta X,\Lie_t\widetilde{\delta X}]
\ee
where $\Lie_t\widetilde{\delta X}$ denotes the initial data for the solution obtained by applying $\Lie_t$ to the solution arising from the initial data $\widetilde{\delta X}$. However, in order to obtain a quantity that is useful for analyzing stability, it is necessary to further restrict the space \(\ms P\) on which $\ms E$ acts in order that the canonical energy be degenerate precisely on the perturbations to other physically stationary stars. The restrictions we need to impose are \(\delta P_i = 0\), where \(P_i\) are the linear momenta at infinity, and that the perturbations be \emph{symplectically-orthogonal} to all the trivial displacements \autoref{eq:triv-gen}. The condition \(\delta P_i = 0\) eliminates the freedom to apply infinitesimal asymptotic boosts to the background solution; such a perturbation makes no physical change but will have non-zero canonical energy.\footnote{\footnotesize Since we do not have a black hole, we do not need to impose the additional horizon conditions of \cite{HW-stab, PW}.\medskip} Symplectic-orthogonality to the trivials makes $\ms E$ degenerate precisely on perturbations to other physically stationary stars (see \cite{GSW} for details).

Following the strategy of \cite{GSW} we impose the constraints  \autoref{eq:lin-constraints} together with the above additional conditions by the following procedure. Consider the space \(\ms W_c \subseteq \ms P\) defined as follows
\be\label{eq:Wc-defn}\begin{split}
	\ms W_c \defn \{& \text{\emph{all} trivial perturbations, and all perturbations produced by} \\ &\text{diffeos that asymptotically approach a spatial translation at infinity}  \}.
\end{split}\ee
Let \(\ms V_c \) denote the subspace of \(\ms P\) which is \emph{symplectically-orthogonal} to \(\ms W_c\),
\be
	\ms V_c \defn {\ms W_c}^{S\perp} = \{ \delta X \in \ms P ~|~ W[\delta X, \widetilde{\delta X}] = 0 \text{ for all } \widetilde{\delta X} \in \ms W_c \}.
\ee
Then any perturbation in \(\ms V_c\) satisfies the constraints, is symplectically-orthogonal to the trivial perturbations and has \(\delta P_i = 0\). Note that the condition \(\delta P_i = 0\) is not a physical restriction as it can be imposed by a suitable asymptotic boost at infinity. As discussed in \cite{GSW}, in the case of axisymmetric perturbations of a stationary-axisymmetric star, symplectic-orthogonality to the (non-flowline) trivials does impose physical restrictions on the perturbations. In particular, since $\eta^\mu = f \phi^\mu$ is a trivial displacement for any function $f$ satisfying $\Lie_u f = \Lie_\phi f = 0$, symplectic-orthogonality to trivial displacements of this form requires the Lagrangian perturbation of the angular momentum density to vanish (which, in particular, requires the total angular momentum of the star to remain unperturbed).

\begin{remark}\label{rem:2-t-derivative}
In the case of stationary-axisymmetric stars with circular flow, the second time derivative of any axisymmetric perturbation \emph{not necessarily obtained from a Lagrangian displacement} is necessarily represented in \(\ms V_c\) (see Lemma 5.1.~of \cite{GSW}). In the case of static stars, following similar arguments, the second time derivative of \emph{any} perturbation not necessarily obtained from a Lagrangian displacement is necessarily represented in $\ms V_c$. Thus for any perturbation of a static star and for axisymmetric perturbations of a stationary-axisymmetric star with circular flow, positivity of $\ms E$ on the space \(\ms V_c\) implies mode stability (see Theorem 5.2.~of \cite{GSW}).
\end{remark} 

	There is significant physical redundancy in \(\ms V_c\), as both infinitesimal diffeomorphisms and trivial perturbations that are symplectically-orthogonal to \(\ms W_c\) are represented in $\ms V_c$. We wish to eliminate this redundancy. For vacuum black hole perturbations, this was done in \cite{PW} by making a concrete gauge choice, as follows: For the vacuum gravitational perturbations, \(p_{ab}\) and \(q_{ab}\) are canonically conjugate variables, and we can thereby define a natural \(L^2\)-inner product for which the symplectic product takes an extremely simple form. We may then fix the gauge completely by demanding \(L^2\)-orthogonality to pure-gauge perturbations. However, for the fluid star, it is much more convenient to perform computations with the variables \(v^a\) and \(\xi^a\). These variables are not canonically conjugate, so there is no corresponding natural \(L^2\)-inner product. Nevertheless, instead of proceeding by fixing all gauge and trivial freedom, we can proceed by simply factoring out the infinitesimal diffeomorphisms and trivial perturbations from the space of perturbed initial data. Define \(\ms W_g\) by
\be\label{eq:Wg-defn}\begin{split}
	\ms W_g \defn \{& \text{\emph{all} trivial perturbations and smooth diffeos in $\ms V_c$ that} \\ &\text{asymptotically approach a translation or rotation at infinity} \}
\end{split}\ee
As shown in \cite{GSW}, the canonical energy \(\ms E\) restricted to \(\ms V_c\) is degenerate precisely on physically stationary perturbations. Since, all the perturbations in \(\ms W_g\) are physically stationary, the canonical energy is also degenerate on \(\ms W_g\). 

	For the stability analysis the perturbations of interest will be in the space
\be\label{eq:V-defn}
	\ms V \defn \ms V_c/\ms W_g.
\ee
It follows that the canonical energy $\ms E$ is well defined on \(\ms V\). In the remainder of this paper we will analyze dynamical stability on \(\ms V\). We will show that if the canonical energy can be made negative on an element of \(\ms V\), then there exist exponentially growing perturbations in the sense that gauge invariant quantities constructed from the perturbation (which are well defined on $\ms V$) grow exponentially.


\section{Positivity of Kinetic Energy}\label{sec:positive-KE}

We now use the $t$ (static) or $t$-$\phi$ (stationary-axisymmetric with circular flow) reflection isometry, $i$, of the background solution to decompose a perturbation into its ``odd'' and ``even'' parts, $P$ and $Q$, under the action of $i$. If the background is static, we consider an arbitrary (smooth, asymptotically flat) perturbation, but if the background is stationary-axisymmetric but non-static, we restrict consideration to axisymmetric perturbations. Let $\Sigma$ be a reflection symmetric Cauchy surface (see \autoref{sec:bg}), with initial data for the perturbation of the form \autoref{eq:init-data-V}. Following \cite{PW} we decompose the space of initial data, $\ms P$, into parts \(\ms P = \ms P_{\rm odd} \oplus \ms P_{\rm even} \) as follows. If the background is static, then $i$ is purely a $t$-reflection, and the \(t\)-reflection odd and even parts of a perturbation, respectively, are given by
\be\label{eq:t-decomp}\begin{split}
	\ms P_{\rm odd} \ni P & \defn \lb( p_{ab}, 0,  v^a, 0 \rb)  \\
	\ms P_{\rm even} \ni Q & \defn \lb( 0, q_{ab},0 , \xi^a \rb) .
\end{split}\ee
In the stationary-axisymmetric case, we first decompose axisymmetric initial data into their ``axial" and ``polar" parts with respect to the axial Killing field \(\phi^a\) as follows:\footnote{\footnotesize This is a local decomposition into the parts that are parallel and orthogonal to the axial Killing field.\medskip}
\be\begin{split}
	p_{ab} & = 2\lambda_{(a}\phi_{b)} + \beta_{ab} + \gamma \phi_{a}\phi_{b} \\
	q_{ab} & = 2\alpha_{(a}\phi_{b)} + \mu_{ab} + \nu \phi_{a}\phi_{b}\\
      v^a &= v_{\parallel} \phi^a + v_{\perp}^a \\
      \xi^a &= \xi_{\parallel} \phi^a +\xi_{\perp}^a
\end{split}\ee
with \(\alpha_a\phi^a = 0 = \lambda_a\phi^a\); $\beta_{ab} = \beta_{(ab)}$, $\mu_{ab} = \mu_{(ab)}$; $\beta_{ab}\phi^a = 0 = \mu_{ab}\phi^a$; $v_{\perp}^a\phi_a=0=\xi_{\perp}^a \phi_a$. Then the ($t$-$\phi$)-reflection odd and even parts, respectively, of an arbitrary axisymmetric perturbation are\footnote{\footnotesize \cite{PW} have a minus sign in front of the \(\alpha_a\) in the definition of \(P\) to maintain canonically conjugate variables. Since, we are not concerned with canonically conjugate variables we omit this sign.\medskip}
\be\label{eq:tphi-decomp}\begin{split}
	\ms P_{\rm odd} \ni P & \defn \lb( \beta_{ab} + \gamma \phi_{a}\phi_{b},~ 2\alpha_{(a}\phi_{b)}, v_{\perp}^a, \xi_{\parallel} \phi^a \rb) \\
	\ms P_{\rm even} \ni Q & \defn \lb( 2\lambda_{(a}\phi_{b)},~ \mu_{ab} + \nu \phi_{a}\phi_{b}, v_{\parallel} \phi^a, \xi_{\perp}^a \rb) .
\end{split}\ee

Since the conditions defining $\ms V_c$ (i.e., $\delta P_i = 0$, symplectic orthogonality to the trivials, and the linearized constraint equations) are invariant under $i$, they cannot couple $P$ and $Q$. Thus, if $(P,Q)$ is a perturbation in $\ms V_c$ then $(P,0)$ and $(0,Q)$ also are in $\ms V_c$, so we similarly have
the decomposition \(\ms V_c = \ms V_{c,\rm odd} \oplus \ms V_{c,\rm even} \). Passing to the space of equivalence classes under $\ms W_g$, we obtain \(\ms V = \ms V_{\rm odd} \oplus \ms V_{\rm even} \). Now, the canonical energy $\ms E$ is constructed from the background spacetime, so it is invariant under $i$ in the sense that for any perturbations $\widetilde{\delta X}$ and $\delta X$ in \(\ms V\), we have ${\ms E} (i^* \delta X, i^* \widetilde{\delta X}) = {\ms E} (\delta X, \widetilde{\delta X})$. It follows that under the decomposition \(\ms V = \ms V_{\rm odd} \oplus \ms V_{\rm even}\), \(\ms E\) cannot contain any ($P$-$Q$)-cross-terms. Thus, $\ms E$ splits up into two quadratic forms $\ms K: \ms V_{\rm odd} \times \ms V_{\rm odd} \to {\mathbb R}$ and $\ms U: \ms V_{\rm even} \times \ms V_{\rm even} \to {\mathbb R}$ such that
\be
\ms E[(\tilde{P}, \tilde{Q}), (P,Q)] = {\ms K} (\tilde{P}, P) + {\ms U} (\tilde{Q}, Q) \, ,
\ee
where
\begin{subequations}\begin{align}
	\begin{split}\label{KE}
		\ms K (\tilde{P}, P) = \ms E[(\tilde{P}, 0), (P,0)]
	\end{split} \\
	\begin{split}\label{PE}
{\ms U} (\tilde{Q}, Q) = \ms E[(0, \tilde{Q}), (0,Q)] \, .
	\end{split}
\end{align}\end{subequations}
We refer to $\ms K(P,P)$ and $\ms U(Q,Q)$, respectively, as the \emph{kinetic energy} and \emph{potential energy} of the perturbation \(\delta X = (P,Q)\). 

We now shall prove that the kinetic energy, $\ms K$, is positive definite on $\ms V_{\rm odd}$.
The analog of this result for perturbations of static/stationary-axisymmetric black hole was proven in 
Theorem 1 of \cite{PW}. To proceed, we need an explicit expression for $\ms K$. 
We cannot directly use the expression given in \cite{HW-stab, PW} as we now have matter fields in the background as well as perturbed matter fields. We shall therefore compute $\ms K$ directly from the definition of canonical energy \autoref{eq:canonical-energy} along with the linearized time-evolution equations.

Since the background is stationary, we have \(\dot p^{ab} = \delta(\tfrac{1}{\sqh}\dot \pi^{ab})\) and \(\dot q_{ab} = \delta(\dot h_{ab})\) and a lengthy computation gives
\begin{subequations}\label{eq:evol-pq}\begin{align}
	\begin{split}\label{eq:evol-q}
	\dot q_{ab} & = \hat\delta h_{ab} + 2N\lb( p_{ab}-\tfrac{1}{2}{p_c}^c~ h_{ab} \rb) - \tfrac{1}{\sqh} N{q_c}^c \lb( \pi_{ab} - \tfrac{1}{2}{\pi_c}^c~ h_{ab} \rb) \\
				&\quad + \tfrac{2}{\sqh}N\lb( 2 q_{c(a}\pi^c_{b)} - \tfrac{1}{2}q_{cd}\pi^{cd} h_{ab} - \tfrac{1}{2}{\pi_c}^c~ q_{ab} \rb) + \Lie_N q_{ab}
	\end{split} \\[1.5ex]
	\begin{split} \label{eq:evol-p}
	\dot p_{ab} & = h_{ac}h_{bd}\tfrac{1}{\sqh}\hat\delta \pi^{cd} +\tfrac{1}{2} N \lb( \Lap q_{ab} + D_aD_b{q_c}^c \rb) - \tfrac{1}{2}D^cN \lb( D_aq_{bc} + D_bq_{ac} - D_cq_{ab} + h_{ab} D_c{q_d}^d \rb)\\
		& \quad + N\lb( 2{R^c}_{(a}q_{b)c} - \tfrac{1}{2}h_{ab}R^{cd}q_{cd} \rb) + h_{ab} D^cND^dq_{cd} - N D^cD_{(a}q_{b)c} + \lb( \Lap N -\tfrac{1}{2}NR\rb)q_{ab} \\
		&\quad  + \tfrac{1}{2} Nh_{ab}\lb( - \Lap {q_c}^c + D^cD^dq_{cd} \rb) - 2 q^c_{(a}D_{b)}D_c N + h_{ab}q^{cd}D_cD_dN - \tfrac{1}{2h}N h_{ab} {q_c}^c \lb( \pi^{cd}\pi_{cd} - \tfrac{1}{2}({\pi_c}^c)^2 \rb)\\
		& \quad + \tfrac{2}{h}N {q_c}^c \lb( \pi_{ac}\pi^c_b - \tfrac{1}{2}{\pi_c}^c\pi_{ab} \rb) - \tfrac{2}{h}N \lb( q_{cd}\pi^c_a\pi^d_b -\tfrac{1}{2} q_{cd}\pi^{cd}\pi_{ab} \rb) - \tfrac{1}{2 h}N q_{ab} \lb( \pi^{cd}\pi_{cd} - \tfrac{1}{2}({\pi_c}^c)^2 \rb) \\
		& \quad -\tfrac{2}{\sqh}N\lb[ 2p_{c(a}\pi^c_{b)} - \tfrac{1}{2}\lb( {p_c}^c\pi_{ab} + {\pi_c}^c p_{ab} \rb) \rb] -\tfrac{1}{2\sqh} {q_c}^c \lb[ D_c\lb( N^c\pi_{ab} \rb) - 2\pi_{c(a}D^cN_{b)}  \rb] \\
		& \quad + D_c\lb( N^c p_{ab} \rb) - 2 p_{c(a}D^cN_{b)} + N h_{ab}\lb[ \tfrac{1}{\sqh}\lb( \pi^{cd} - \tfrac{1}{2}{\pi_e}^e h^{cd} \rb)p_{cd} + \tfrac{1}{h}\lb( \pi^{ce} {\pi_e}^d - \tfrac{1}{2} {\pi_e}^e\pi^{cd} \rb) q_{cd} \rb] \\
		& \quad  + 8 \pi N \tau_{ab}
	\end{split}
\end{align}\end{subequations}
where \(\tau^{ab} = \delta T^{ab}\). The evolution equations for the perturbed fluid initial data are obtained in Appendix~\ref{fluideqs}; they are
\begin{subequations}\label{eq:evol-xiv}\begin{align}
\begin{split}\label{eq:dotxi}
	\dot\xi^a & = \hat\xi^a+ \Lie_N \xi^a - \tfrac{N}{\sqrt{1+u^2}} \Lie_u \xi^a\\
	&\quad +\tfrac{N}{\sqrt{1+u^2}}\left({\delta^a}_d -\tfrac{1}{1+u^2} u^a u_d\right) \left[v^d - \tfrac{1}{2} u^d u^b u^c q_{bc} -N\sqrt{1+u^2}\, u^d  \xi^b D_b \left(\tfrac{\sqrt{1+u^2}}{N} \right) \right]
	\end{split} \\[1.5ex]
	\begin{split} \label{eq:dotv}
 \dot v^a & = \hat v^a + \left({\delta^a}_e + \tfrac{c_s^2}{1+(1-c_s^2)u^2} u^a u_e\right) \times \Biggl\{
   \Lie_N v^e - \tfrac{1}{2\sqrt{1+u^2}}\left(q_{bc}u^b u^c + 2v^b u_b\right)D^e N  \\
	&\quad + \sqrt{1+u^2}\,q^{eb} D_b N  +N\left(u^e{p^b}_b  -2 u_b p^{eb}\right)- \tfrac{N}{2\sqrt{1+u^2}}u^b u^c  \left(2 D_b {q_c}^e - D^e q_{bc}  \right) \\
	&\quad  +\tfrac{N}{\sqrt{h}}\left(  \pi^{eb} u_b{q^c}_c -2\pi^{eb}u^c q_{bc}
 -\tfrac{1}{2} {\pi^b}_b u^e{q^c}_c   + \pi^{bc} u^e q_{bc}+{\pi^b}_b v^e -2 \pi^{eb}v_b\right) \\
	&\quad +\tfrac{N}{2(1+u^2)^{3/2}}(q_{cd}u^c u^d + 2v^c u_c)u^b D_b u^e - \tfrac{N}{\sqrt{1+u^2}}(v^b D_b u^e+ u^b D_b v^e) \\
	&\quad - \frac{1+c_s^2}{\rho+p}  \left(-\frac{\Delta n}{n}  + \frac{\xi^c D_c \ln(\rho+p)}{1+c_s^2}\right) \left( \tfrac{N}{\sqrt{1+u^2}}\left(h^{eb}+ u^e u^b \right) -u^e N^b \right) D_bp \\
	&\quad  +\frac{N(q_{bc}u^b u^c + 2v^b u_b)\left(D^e p+ u^e u^b D_b p\right)}{2(\rho+p)(1+u^2)^{3/2}} + \frac{v^e  N^b D_b p}{\rho+p}  +\frac{N\left( q^{eb} D_b p- 2 v^{(e} u^{b)}D_b p\right)}{(\rho+p)\sqrt{1+u^2}}  \\ 
	&\quad - \left(\frac{\sqrt{1+u^2}\,u^e N^b-N\left(h^{eb} +u^e u^b \right)}{(\rho+p)\sqrt{1+u^2}}\right) D_b \left[- c_s^2 (\rho+p) \frac{\Delta n}{n} +\xi^c D_c p\right]\\
	&\quad   +\frac{1}{2}c_s^2 u^e\left(h^{bc} + \frac{u^b u^c}{1+u^2}\right)\dot q_{bc} + \frac{c_s^2 u^e D_b\left(\sqrt{1+u^2}\, \dot\xi^b\right)}{\sqrt{1+u^2}}  + \frac{u^e \dot \xi^b D_b p}{(\rho+p)}   \Biggr\}
	\end{split}
\end{align}\end{subequations}
where 
\be\begin{aligned}\label{Deltanovern}
\frac{\Delta n}{n} &= -\frac{1}{2}\left(h^{ab} + \frac{u^a u^b}{1+u^2}\right)q_{ab} - \frac{ u_a v^a}{1+u^2}-\frac{D_a\left(\sqrt{1+u^2}\, \xi^a\right)}{\sqrt{1+u^2}} 
\end{aligned}\ee
and one must substitute \autoref{eq:evol-q} and \autoref{eq:dotxi} in the last line of \autoref{eq:dotv} to get an explicit expression in terms of initial data.

The quantities the \(\hat p_{ab}\), \(\hat q^{ab}\), $\hat \xi^a$ and $\hat v^a$ in \autoref{eq:evol-pq}-\autoref{eq:evol-xiv} are pure gauge perturbations generated by diffeomorphisms due to the perturbed lapse and shift. Since the canonical energy is gauge-invariant, we will ignore these terms in our computation of the kinetic energy below, i.e., we will perform the calculations assuming the perturbed lapse and shift are zero.\\

The kinetic energy is obtained by substituting a reflection-odd perturbation into the definition of the canonical energy \autoref{eq:canonical-energy}
\be\label{eq:KE-master}\begin{split}
	\ms K & = \ms K_{\rm GR} + \ms K_{\rm fluid} \\
		& = \frac{1}{16\pi}\int \df\varepsilon^{(3)}\lb( p_{ab} \dot q^{ab} - q^{ab} \dot p_{ab} \rb) + \int \left(\dot \xi^\mu\delta P_{\mu\nu\lambda\rho}  -  \xi^\mu {\dot{(\delta P)}}_{\mu\nu\lambda\rho} - [\xi,\dot\xi]^\mu P_{\mu\nu\lambda\rho}\right).
\end{split}\ee\\

We first compute the kinetic energy in the simpler case of a static star where the reflection-odd perturbation is given by \(\delta X = \lb( p_{ab}, 0,  v^a, 0 \rb)\). Using \autoref{eq:evol-q} for a static spacetime background the first term in \autoref{eq:KE-master} becomes
\be
	\ms K_{\rm GR} = \frac{1}{16\pi}\int \df\varepsilon^{(3)} p_{ab} \dot q^{ab} = \frac{1}{8\pi}\int \df\varepsilon^{(3)} N\lb( p_{ab}p^{ab}-\tfrac{1}{2}({p_a}^a)^2 \rb).
\ee
To compute the second term in \autoref{eq:KE-master}, note that \autoref{eq:dotxi} gives
\be
\dot\xi^a = N v^a,
\ee
since $u^a=0$ for the static background.
Furthermore we obtain
\be
\delta u^\mu = \delta\left(\sqrt{1+u^2} \nu^\mu + {h^\mu}_a u^a\right) = \delta({h^\mu}_a u^a) = {h^\mu}_a v^a
\ee
where $\delta \nu^\mu = 0$ follows from the vanishing of the perturbed lapse and shift. Thus, the pullback to $\Sigma$ of $\dot \xi^\mu \delta P_{\mu\nu\lambda\rho}$ is
\be\begin{split} \label{statxideltaP}
\dot \xi^\mu \delta P_{\mu\nu\lambda\rho} \Bigr|_\Sigma &= N v^\mu \delta \left[ (\rho+p) ({\delta_\mu}^\sigma + u_\mu u^\sigma)\varepsilon_{\sigma\nu\lambda\rho} \right] \Bigr|_\Sigma = N v^\mu (\rho+p) \varepsilon_{\sigma\nu\lambda\rho}\delta \left(  g_{\mu \chi} u^\chi u^\sigma \right) \Bigr|_\Sigma\\
&= N (\rho+p)v^a v_a \varepsilon^{(3)}_{\nu\lambda\rho} 
\end{split}\ee
where we have used the fact that the pullback to $\Sigma$ of ${h_\mu}^\sigma \varepsilon_{\sigma\nu\lambda\rho}$ vanishes, and in the last equality we used the fact that the vanishing of the perturbed lapse and shift imply that \(\delta g_{\mu\nu}\) is tangent to \(\Sigma\) and vanishes (since \(q_{ab}=0\)). 
The other parts of the second term in \autoref{eq:KE-master} vanish by $\xi^a=0$, and
thus for the static star the kinetic energy is
\be\label{eq:KE-fluid-static}
	\ms K = \frac{1}{8\pi}\int \df\varepsilon^{(3)} N\lb( p_{ab}p^{ab}-\tfrac{1}{2}({p_a}^a)^2 \rb) + \int \df\varepsilon^{(3)} N (\rho+p) v_a v^a
\ee\\

Now we generalize the above computation for the case of a stationary-axisymmetric star with axisymmetric perturbations. The reflection-odd initial data is now given by \(\delta X = \lb( \beta_{ab} + \gamma \phi_{a}\phi_{b},~  2\alpha_{(a}\phi_{b)}, v^a, \xi \phi^a \rb) \) with \(v^a \phi_a = 0 \).  
However, a perturbation of the form $\lb( 0, 0, 0, \xi \phi^a \rb)$ is a trivial and a degeneracy of the canonical energy and so will not contribute to our calculation of the kinetic energy. Thus, without loss of generality, we may take the reflection-odd initial data to be \(\delta X = \lb( \beta_{ab} + \gamma \phi_{a}\phi_{b},~  2\alpha_{(a}\phi_{b)}, v^a, 0 \rb)\) with $v^a \phi_a = 0$. Again we set the perturbed lapse and shift to zero, without loss of generality. Note that, for reflection-odd initial data \(q_{ab}\) is traceless and \(\tau^{ab} = \delta T^{ab} = 2 \tau^{(a}\phi^{b)}\) i.e. \(\tau^{ab}\) is axial.

	We now compute the first term of \autoref{eq:KE-master}, using the linearized evolution equations \autoref{eq:evol-q} and \autoref{eq:evol-p}. The resulting expression is simplified by using the following steps:
\begin{enumerate}
	\item ``Integrate by parts'' any term with two derivatives of \(q_{ab}\) to rewrite it as a quadratic expression in one derivative of \(q_{ab}\).
	\item Write the shift vector as \(N^a = \bar N \phi^a\) and use the axisymmetry of the perturbations and \autoref{eq:shift-id} to write the terms in terms of \(\pi^{ab}\) and \(q_{ab}\).
	\item Rewrite the terms with Ricci tensor, Ricci scalar and two derivatives of the lapse \(N\) using \autoref{eq:constraints-H-ax}, \autoref{eq:Lap-N-id} and \autoref{eq:Ric-id} in terms of the background matter terms.
\end{enumerate}

Thus the relevant contributions to the first term of \autoref{eq:KE-master} are (in the following intermediate expressions we have omitted the spatial volume element \(\df\varepsilon^{(3)}\))
\be\begin{split}
	16\pi \ms K_{\rm GR} & =  2\int N\lb[ p_{ab}p^{ab}-\tfrac{1}{2}({p_c}^c + \tfrac{1}{\sqh}q_{ab}\pi^{ab})^2\rb] + 8\int N \lb[ \Phi p^{ab} \lb( \alpha_a\pi_{b} \rb) + \Phi^2\gamma \alpha_c\pi^{c}\rb] \\ 
	& \quad +  4\int N \Phi^2(\alpha_a \pi^{a}) (\alpha^b \pi_{b}) -\int 4 N \Phi p_{ab}\alpha^a \pi^{b} + \int N \lb[\tfrac{1}{2} (D_cq_{ab})^2 - D_cq_{ab} D^aq^{bc} \rb] \\
	& \quad -4 \int N \Phi^2\lb[ (\pi_a \pi^{a}) (\alpha^b \alpha_{b}) - (\alpha_a \pi^{a}) (\alpha^b \pi_{b})\rb] -16\pi \int N\left[\Phi(\alpha^a \alpha^{b})\hat V_{ab} + \Phi^2(\alpha^{a}\alpha_a)\bar V \right]\\
	& \quad + 8\pi\int N \Phi\lb( E - T_c^c \rb)(\alpha_a \alpha^a) -4\int N \Phi p_{ab} \alpha^a\pi^{b} - 16\pi\int N \Phi\alpha^a \tau_a
\end{split}\ee
where since \(V^{ab}\) is polar we have written \(V^{ab} = \hat V^{ab} + \bar V \phi^a \phi^b\) with \(\hat V^{ab}\phi_a = 0\).

We can further write \(\int N \lb[\tfrac{1}{2} (D_cq_{ab})^2 - D_cq_{ab} D^aq^{bc} \rb]\) in the form
\be\label{K1}
	\ms K_1 =2\int N \Phi^{-1}\lb[D_{[a}(\Phi\alpha_{b]})D^{[a}(\Phi\alpha^{b]}) - \frac{1}{2} (\alpha^{a}D_a\Phi)(\alpha^{b}D_b\Phi) + (\alpha^{a} D^b\Phi )D_a(\Phi\alpha_{b}) \rb].
\ee
The last term of \autoref{K1} can be written as
\be
	2\int N \Phi^{-1}(\alpha^{a} D^b\Phi )D_a(\Phi\alpha_{b}) = 2\int N\Phi^{-1} \lb[ (\alpha^{a} D^b\Phi )D_b(\Phi\alpha_{a}) + 2(\alpha^{a} D^b\Phi)D_{[a}(\Phi\alpha_{b]}) \rb].
\label{K13}
\ee
This expression can be simplified by using the relation
\be\label{eq:Ric-Killing}
R_{ab}\phi^b = -\tfrac{1}{2} \phi_{a} D^b\lb(  \Phi^{-1} D_b\Phi \rb)
\ee
(which holds by virtue  of $\phi^a$ being a Killing field) and then contracting again with the Killing field eliminating $R_{ab}$ using the background ADM equation \autoref{eq:Ric-id} to obtain
\be
	D^a\lb( N\Phi^{-1}D_a\Phi \rb) = -4N\Phi\pi_a\pi^a - 16\pi N\Phi \bar V \, .
\ee
Using this relation, we simplify \autoref{K13} as follows:
\be\begin{split}
		& 2\int N\Phi^{-1} (\alpha^{a} D^b\Phi )D_b(\Phi\alpha_{a}) = -2 \int \lb[ D^b\lb( N\Phi^{-1}D_b \Phi\rb)\Phi\alpha^{a}\alpha_{a} + N \alpha_a D^b\alpha^{a} D_b\Phi \rb]\\
		& = 8\int N \Phi^2\lb(\alpha^{a}\alpha_a\rb)\lb(\pi_{a}\pi^a \rb) - 2\int N\Phi^{-1} (\alpha^{a} D^b\Phi )D_b(\Phi\alpha_{a}) -2\int N\Phi\alpha^{a}\alpha_{a}D_b \Phi^{-1}D^b\Phi \\
		&\quad + 32\pi\int N\Phi^2\lb(\alpha^{a}\alpha_a \rb) \bar V \\
		& = 4\int N\Phi^2 \lb(\alpha^{a}\alpha_a\rb)\lb(\pi_{a}\pi^a \rb) - \int \Phi\alpha^{a}\alpha_{a}D_b \Phi^{-1}D^b\Phi + 16\pi\int N \Phi^2\lb(\alpha^{a}\alpha_a \rb) \bar V.
	\end{split}\ee
Thus, we obtain
	\[\begin{split}
	\ms K_1 & = 2\int N \Phi(D_{[a}\alpha_{b]})(D^{[a}\alpha^{b]}) + 4\int N\Phi^2 \lb(\alpha^{a}\alpha_a \rb)\lb(\pi_{a}\pi^a \rb) + 16\pi\int N\Phi^2\lb(\alpha^{a}\alpha_a \rb) \bar V
	\end{split}\]
and we have
\be\begin{split}
	16\pi \ms K_{\rm GR} & =  2\int N\lb[ \Phi(D_{[a}\alpha_{b]})(D^{[a}\alpha^{b]}) + \beta_{ab}\beta^{ab} + \Phi^2 \lb( \gamma + 2\alpha_a \pi^{a} \rb) \lb( \gamma + 2\alpha^b \pi_{b} \rb) \rb. \\
	&\qquad \lb. - \tfrac{1}{2}\lb({\beta_a}^a + \Phi\gamma + 2\Phi \alpha_a \pi^a \rb)^2 \rb] \\
	&\quad -16\pi \int N\Phi(\alpha^a \alpha^{b})\hat V_{ab} + 8\pi\int N \Phi\lb( E - T_c^c \rb)  (\alpha_a \alpha^a) - 16\pi\int N\Phi \alpha^a \tau_a.
\end{split}\ee
Computing the contribution due to \(\hat V_{ab}\) using \autoref{eq:V-fluid-defn} we get
\be\begin{split}
	\ms K_{\rm GR} & =  \frac{1}{8\pi}\int N\lb[ \Phi(D_{[a}\alpha_{b]})(D^{[a}\alpha^{b]}) + \beta_{ab}\beta^{ab} + \Phi^2\lb( \gamma + 2\alpha_a \pi^{a} \rb)^2 \rb. \\
	&\qquad \lb. - \tfrac{1}{2}\lb({\beta_a}^a + \Phi\gamma + 2\Phi\alpha_a \pi^a\rb)^2  -  8\pi \Phi\alpha^a ( \alpha^{b} T_{ab} + \tau_a ) \rb].
\end{split}\ee
To compute the final term in the above expression we note that the \(t\)-\(\phi\)-reflection-odd fluid perturbation is given by the initial data listed in \autoref{eq:tphi-decomp} i.e. \(\xi^a = 0\) and \(\phi_a v^a = 0\). From the fact that the perturbation is \(t\)-\(\phi\)-reflection-odd it follows immediately that $\delta\rho=\delta p=0$, since axisymmetric scalars must be \(t\)-\(\phi\)-reflection-even. Using \autoref{eq:circular-flow} and \autoref{eq:fluid-matter}, we obtain
\be
\alpha^a \alpha^b T_{ab} = p\alpha_a \alpha^a
\ee

\be\begin{split}
\tau^{ab} = \delta T^{ab} &= \delta\left[(\rho+p)u^a u^b + p h^{ab}\right] = 2(\rho+p) U \phi^{(a} v^{b)} - 2 p \phi^{(a} \alpha^{b)} \\
\implies \tau^a &= (\rho+p)U v^{a} - p \alpha^{a} \\
\text{and}\quad \alpha^a \tau_a & = (\rho+p)U \alpha_a v^{a} - p \alpha_a\alpha^{a}
\end{split}\ee
and thus, 
\be\begin{split}
\alpha^a \alpha^b T_{ab}+\alpha^a \tau_a=  (\rho+p)U \alpha_a v^{a}.
\end{split}\ee

	Next, we calculate the second term of \autoref{eq:KE-master}. From \autoref{eq:dotxi} we find
\be
\dot \xi^a =  \frac{N}{\sqrt{1+\Phi U^2}}v^a.
\ee
The last two terms in \autoref{eq:KE-master} vanish because $\xi^a=0$, and the pullback to $\Sigma$ of $\dot\xi^\mu \delta P_{\mu\nu\lambda\rho}$ is
\be\begin{split}\label{stationaryPcalc}
\dot\xi^\mu \delta P_{\mu\nu\lambda\rho} \Bigr|_\Sigma&= \dot\xi^\mu \delta \left[ (\rho+p) ({\delta_\mu}^\sigma + u_\mu u^\sigma)\varepsilon_{\sigma\nu\lambda\rho} \right]\Bigr|_\Sigma = \dot\xi^\mu (\rho+p)\sqrt{1+\Phi U^2}\, \varepsilon^{(3)}_{\nu\lambda\rho}\delta (u_\mu)\\
&= N v^\mu(\rho+p) \varepsilon^{(3)}_{\nu\lambda\rho}\left( \Phi U \alpha_\mu + v_\mu \right).
\end{split}\ee
Thus, we have
\be\begin{split}
\ms K_{\rm fluid} &= \int (\rho+p)N\left( \Phi U  \alpha_{a}v^a + v_a v^a \right) .
\end{split}\ee

Thus, the total kinetic energy for the stationary-axisymmetric Einstein-perfect fluid star is
\be\label{eq:KE-fluid-station}\begin{split}
\ms K & = \frac{1}{8\pi}\int \df\varepsilon^{(3)}  N\biggl[ \Phi \left(D_{[a}\alpha_{b]}\right)\left(D^{[a}\alpha^{b]}\right) + \beta_{ab}\beta^{ab} + \Phi^2\left(\gamma+2 \alpha_a \pi^a \right)^2 \\
	&\qquad\qquad - \frac{1}{2}\left({\beta_a}^a+\Phi\gamma+2\Phi\alpha_a \pi^a \right)^2 \biggr] + \int \df\varepsilon^{(3)} N(\rho+p) v^a v_a.
\end{split}\ee\\

\begin{thm}[Positivity of kinetic energy]\label{thm:positive-KE}
For arbitrary perturbations of a static Einstein-perfect fluid star, and
	for axisymmetric perturbations of a stationary-axisymmetric Einstein-perfect fluid star with circular flow, the kinetic energy \(\ms K\) (given by \autoref{eq:KE-fluid-static} for the static case, and \autoref{eq:KE-fluid-station} for the stationary-axisymmetric case) is a positive definite symmetric bilinear form on \(\ms V_{\rm odd}\).
\begin{proof}
The proof closely parallels the vacuum case given in Theorem 1 of \cite{PW}. We start first with the static case where the kinetic energy is given by a simpler expression \autoref{eq:KE-fluid-static}. Let \(f\) be a solution to the boundary value problem\footnote{\footnotesize Note that the corresponding Eq.~4.21 (and Eq.~4.25 for the stationary-axisymmetric case) in \cite{PW} is missing a negative sign on the right-hand-side of the elliptic equation and a factor of \(1/r\) in the asymptotic conditions at infinity.\medskip}
	\be\label{eq:f-eqn}
		\lb[ D^2 - 4 \pi (\rho + 3p) \rb] f = -\frac{1}{2}p_c{}^c \quad\text{with}\quad f\vert_\infty \sim O(1/r).
	\ee
	By \autoref{fluidconditions}, the elliptic operator on the left-hand-side is negative\footnote{\footnotesize We need the strong energy condition, $(\rho+3p)\geq 0$ and $(\rho+p)\geq 0$ to get a unique solution to \autoref{eq:f-eqn} (and \autoref{eq:f-eqn-station}). These follow from the stronger conditions of \autoref{fluidconditions}, $\rho \geq 0$ and $p\geq 0$, which are assumed in showing the well-posedness of the Einstein-perfect fluid initial value problem \cite{CBYork}.\medskip} and thus the above boundary value problem has a unique solution. Define \(\hat{p}_{ab}\) and \(\hat v^a\) by
	\be\label{eq:new-p-v}\begin{split}
	\hat{p}_{ab} & \defn p_{ab} - D_aD_b f + h_{ab}\Lap f + f \lb( R_{ab} - \tfrac{1}{2} R h_{ab}\rb) - 8 \pi f T_{ab} \\
	\hat v^a & \defn v^a + D^a f + \frac{D^ap}{\rho+p} f.
	\end{split}\ee
	A direct computation shows that since \(p_{ab}\) and \(v^a\) satisfy the linearized momentum constraint \autoref{momconstraint}, so do \(\hat p_{ab}\) and \(\hat v^a\), that is 
	\be
	D_b p^{ab} = - 8\pi (\rho+p)v^a \implies D_b \hat p^{ab} = - 8\pi (\rho+p) \hat v^a.
	\ee
	Further, from \autoref{eq:f-eqn}, the background Hamiltonian constraint \autoref{eq:constraint-H}, and \autoref{eq:fluid-matter} we get \(\hat p_c{}^c = 0 \). We use \autoref{eq:new-p-v} to replace \(p_{ab}\) and \(v^a\) in the kinetic energy \autoref{eq:KE-fluid-static} in favor of \(\hat p_{ab}\) and \(\hat v^a\). Integrating by parts, we find that the \(f\)-\(f\), \(f\)-\(\hat p\) and \(f\)-\(\hat v\) terms vanish. Thus, we obtain
	\be
		\ms K = \frac{1}{8\pi}\int \df\varepsilon^{(3)} N\lb( \hat p_{ab} \hat p^{ab}\rb) + \int \df\varepsilon^{(3)} N (\rho+p) \hat v_a \hat v^a \geq 0.
	\ee
	Thus, \(\ms K\) is manifestly non-negative and vanishes if and only if \(\hat p_{ab} = \hat v^a = 0\). In that case, we see from \autoref{eq:new-p-v} that \(p_{ab}\) and \(v^a\) are pure-gauge perturbations in \(\ms W_g\) (\autoref{eq:Wg-defn}) generated by diffeomorphisms along \(f \nu^\mu\). Since we have factored out such perturbations, we see that \(\ms K\) is positive definite on \(\ms V_{\rm odd}\).

The proof for the stationary-axisymmetric star follows from very similar arguments. Now let \(f\) be a solution to the following boundary value problem
	\be\begin{aligned}\label{eq:f-eqn-station}
		\lb[ D^2 - 2\Phi \pi_a \pi^a - 4 \pi \lb( \rho + 3p + (\rho + p)\Phi U^2 \rb) \rb]f &= - \frac{1}{2} \lb( {\beta_a}^a+\Phi\gamma+2\Phi\alpha_a\pi^a \rb) \\
 \quad\text{with}\quad f\vert_\infty &\sim O(1/r)
	\end{aligned}\ee
	where we have used \(\pi^{ab} = \sqh \pi^{(a}\phi^{b)}\) and \(u^a = U \phi^a\) in the background. Again, \autoref{fluidconditions} ensures that \autoref{eq:f-eqn-station} has a unique solution. Define now
	\be\label{eq:new-p-v-station}\begin{split}
	\hat{p}_{ab} & \defn p_{ab} - D_aD_b f + h_{ab}\Lap f + f \lb( R_{ab} - \tfrac{1}{2} R h_{ab}\rb) - f h_{ab} (2\Phi \pi_c \pi^c) \\ 
		&~~ \quad  + 2 f (\Phi \pi_a \pi_b + \pi_c\pi^c \phi_a \phi_b)  - 8 \pi f T_{ab} \\
	\hat{\alpha}_a & \defn \alpha_a - 2f \pi_a \\
	\hat v^a & \defn v^a + \sqrt{1+\Phi U^2} D^a f + \lb[ \frac{D^a p}{(\rho+p)\sqrt{1+\Phi U^2}} + 2U \pi^a - \frac{U^2 D^a \Phi}{2\sqrt{1+\Phi U^2}} \rb] f . \\
	\end{split}\ee
	Note that \(\hat \alpha_a\) and \(\hat v^a\) are axial while \(\hat p_{ab} = \hat \beta_{ab} + \hat \gamma \phi_a\phi_b\) is polar. The new variables satisfy the linearized momentum constraint \autoref{momconstraint}, i.e.,
	\be\begin{split}
		D_b\beta^{ab} -\tfrac{1}{2}\gamma D^a \Phi +\pi_{b}\lb( 2\Phi D^{[b}\alpha^{a]}-\alpha^b D^a \Phi \rb) & = - 8\pi (\rho+p)\sqrt{1+\Phi U^2}~ v^a \\
	\implies D_b\hat\beta^{ab} -\tfrac{1}{2}\hat \gamma D^a \Phi +\pi_{b}\lb( 2\Phi D^{[b}\hat\alpha^{a]}-\hat\alpha^b D^a \Phi \rb) & = - 8\pi (\rho+p)\sqrt{1+\Phi U^2}~ \hat v^a.
	\end{split}\ee

 Further, using \autoref{eq:f-eqn-station}, \autoref{eq:constraint-H} and \autoref{eq:fluid-matter}, we get \({\hat\beta_a}{}^a+\Phi\hat\gamma + 2\Phi\hat\alpha_a\pi^a = 0\), and in parallel to the arguments in the static case we can write the kinetic energy in a manifestly non-negative form
	\be\begin{split}
\ms K & = \frac{1}{8\pi}\int \df\varepsilon^{(3)}  N \lb[ \Phi \left(D_{[a}\hat\alpha_{b]}\right)\left(D^{[a}\hat\alpha^{b]}\right) + \hat\beta_{ab}\hat\beta^{ab} + (\Phi\hat\gamma + 2\Phi\hat\alpha_a\pi^a )^2 \rb] \\
	&~ \quad + \int \df\varepsilon^{(3)} N(\rho+p) \hat v^a \hat v_a \geq 0.
\end{split}\ee
	The kinetic energy vanishes if and only if we have
	\be
	\hat \alpha_a = D_a \hat f \eqsp \hat\gamma = - 2 \pi^a D_a \hat f \eqsp \hat \beta_{ab} = \hat v^a = 0
	\ee
	for some \(\hat f\), in which case \(p_{ab}\), \(\alpha_a\) and \(v^a\) in \autoref{eq:new-p-v-station} are pure-gauge perturbations in \(\ms W_g\) generated by a diffeomorphism along \(f \nu^\mu + \hat f \phi^\mu\). Again, since we have factored out such perturbations, \(\ms K\) is positive definite on \(\ms V_{\rm odd}\).
\end{proof}
\end{thm}

	The transformations in \autoref{eq:new-p-v} and \autoref{eq:new-p-v-station} are gauge transformations corresponding to making a normal displacement of the Cauchy surface \(\Sigma\) by \(f\). The condition \({\hat\beta_a}{}^a+\Phi\hat\gamma + 2\Phi\hat\alpha_a\pi^a = 0\) is simply the condition that \(\hat \delta \pi_a{}^a = 0\) and thus, writing the kinetic energy in terms of \(\hat p_{ab}\), \(\hat\alpha_a\) and \(\hat v^a\) corresponds to working in a gauge where \(\Sigma\) is a maximal slice in the perturbed spacetime.

\section{Negative Energy and Exponential Growth}\label{sec:exp-growth}

Consider a smooth, axisymmetric perturbation \(\delta X \in \ms V_c\). The time evolution of $\delta X$ is given by Eqs.\ref{eq:evol-pq}-\ref{eq:evol-xiv} of the previous section. These equations contain arbitrary gauge transformations on the right-hand-side. However, we may effectively remove this gauge dependence by simply factoring out by the space \(\ms W_g\) (see \autoref{eq:Wg-defn}) to the pass to the space $\ms V$ defined by \autoref{eq:V-defn}. By a slight abuse of notation, we will continue to denote by $\delta X$ the element of $\ms V$ corresponding to the equivalence class of the original perturbation $\delta X \in \ms V_c$. Similarly, we will write $\dot{\delta X}$ for the equivalence class of $\Lie_t \delta X$. The time evolution equations \autoref{eq:evol-pq}-\autoref{eq:evol-xiv} then can be written using an operator $\mc E:\ms V \to \ms V$ as
 as
\be\label{eq:evol-op-defn}
	\dot{\delta X}  = \mc E (\delta X)
\ee
The time evolution operator $\mc E$ is related to the canonical energy $\ms E$ by
\be\label{eq:E-W}
	\ms E(\widetilde{\delta X}, \delta X) = W(\widetilde{\delta X}, \mc E(\delta X))
\ee
which expresses the fact that the canonical energy is a Hamiltonian for the time evolution of the linearized perturbations. 

We now make use of the reflection isometry $i$ of the background spacetime. As in the previous section, we decompose initial data $\delta X \in \ms V$ into its odd part, $P \in \ms V_{\rm odd}$, and even part, $Q \in \ms V_{\rm even} $ under the action of $i$ (see \autoref{eq:tphi-decomp}). Since the time evolution operator $\mc E$ is invariant under $i$, the  evolution equations take the form:
\begin{subequations}\label{eq:evol-odd-even}\begin{align}
	\dot Q & = \mc K P  \label{eq:evol-K} \\
	\dot P & = -\mc U Q \label{eq:evol-U}
\end{align}\end{subequations}
where, the maps $\mc K$ and $\mc U$ act as
\be
	\mc K : \ms V_{\rm odd} \to \ms V_{\rm even} \eqsp \mc U : \ms V_{\rm even} \to \ms V_{\rm odd} \, . \\
\ee
Explicit formulae for \(\mc K\) and \(\mc U\) can be obtained by substitution of an odd or, respectively, even perturbation from \autoref{eq:t-decomp} or \autoref{eq:tphi-decomp} into the right-hand-sides of the evolution equations Eqs.\ref{eq:evol-q}-\ref{eq:dotv}. In particular, in the case of a static background, the odd part of a perturbation is given by $P = ( p_{ab}, 0,  v^a, 0)$, and from Eqs.\ref{eq:evol-q}-\ref{eq:dotv}, it follows that 
\be\label{Kstatic}
\mc K \begin{pmatrix}p_{ab}  \\ 0 \\v^a\\ 0 \end{pmatrix} = \begin{pmatrix} 
		0 \\ 2N\left( p_{ab}-\tfrac{1}{2}{p_c}^c h_{ab} \right) \\
		0 \\ N v^a\end{pmatrix}
\ee
In the case of a stationary, axisymmetric background, $P$ is given by \autoref{eq:tphi-decomp}, and ${\mc K} P$ is given by
\be\label{eq:Kstation}
\mc K \begin{pmatrix}\beta_{ab} +\gamma \phi_a \phi_b\\2 \alpha_{(a}\phi_{b)}\\v^a\\\xi \phi^a\end{pmatrix} = \begin{pmatrix} 2\Lambda_{(a}\phi_{b)} \\ \Theta_{ab} + \Gamma \phi_{a}\phi_{b} \\ V \phi^a \\ \Xi^a \end{pmatrix}
\ee
where we have used the decomposition \autoref{eq:tphi-decomp} for the reflection-even perturbation on the right-hand-side with
\begin{subequations}\label{eq:K-op-exp}\begin{align}
	\Lambda_a & = - \Phi^{-1} D^b \lb( N \Phi D_{[a}\alpha_{b]} \rb) + N \pi_a \lb( \beta_b{}^b - \Phi\gamma - 2\Phi \alpha_b\pi^b \rb) + 8 \pi N (\rho+p)U v_a \\[1.5ex]
	\Theta_{ab} & = 2N\left[ \beta_{ab} - \tfrac{1}{2}({\beta_c}^c+\Phi\gamma+2\Phi\alpha_c \pi^c) ( h_{ab} - \Phi^{-1}\phi_a \phi_b ) \right] \\[1.5ex]
	\Gamma  & = 2N\left[ (\gamma+2\alpha_c \pi^c) -\tfrac{1}{2}\Phi^{-1}({\beta_c}^c+\Phi\gamma+2\Phi\alpha_c \pi^c) \right] \\[1.5ex]
	V  & = \frac{1+\Phi U^2}{1+(1-c_s^2)\Phi U^2} \lb[ N U \lb({\beta_a}^a - \Phi\gamma - 2\Phi \alpha_a\pi^a \rb) - \frac{Nv^a D_a(\Phi U) }{\Phi\sqrt{1+\Phi U^2}} + \frac{U c_s^2D_a(Nv^a)}{\sqrt{1+\Phi U^2}}  \rb. \\
	& ~\quad \lb. - \frac{N U c_s^2 }{2(1+\Phi U^2)}\left[(1+2\Phi U^2) {\beta_a}^a + \Phi\gamma + 2\Phi \alpha_a \pi^a \right]  \rb] \nonumber \\[1.5ex]
	\Xi^a & =  \frac{N}{\sqrt{1+\Phi U^2}}v^a
\end{align}\end{subequations}
where we have used \autoref{eq:shift-id}, \autoref{eq:Ric-id} and \autoref{eq:Ric-Killing} for the background spacetime.

	The formula for \(\mc U\) can be computed in the same manner, substituting $Q$ on the right sides of Eqs.\ref{eq:evol-q}-\ref{eq:dotv} instead of $P$. However, since this formula is considerably more complicated, we will not attempt to write it out explicitly here.

The decomposition of the canonical energy \(\ms E\) into the \emph{kinetic} and \emph{potential} energies is given in terms of these operators by
\be\label{eq:K-U-W}
	\ms K(\tilde P, P) \defn W[\tilde P, \mc K P] \eqsp \ms U(\tilde Q, Q) \defn -W[\tilde Q, \mc U Q].
\ee

Taking the time derivative of \autoref{eq:evol-K} and using \autoref{eq:evol-U}, we obtain
\be\label{ddotQ}
\ddot Q = -\mc A Q 
\ee
where
\be
\mc A \defn \mc K\mc U  : \ms V_{\rm even} \to \ms V_{\rm even}  \, .
\label{Adef}
\ee
Following \cite{PW}, we now define a new Hilbert space that makes $\mc A$ a symmetric operator, thereby allowing us solve this equation by spectral methods.  Let $\mc K[\ms V_{\rm odd}] \subseteq \ms V_{\rm even}$ denote the range of the operator $\mc K$. By \autoref{thm:positive-KE}, \(\mc K\) is a positive definite operator on \(\ms V_{\rm odd}\), so $\mc K$ has vanishing kernel. Thus for all \( Q \in \mc K[\ms V_{\rm odd}] \), there exists a unique \( P \in \ms V_{\rm odd} \) such that, \( Q = \mc K P \). Using this fact, we define a new inner product, $\inp{\phantom{\_},\phantom{\_} }_{\ms H}$ on \( \mc K[\ms V_{\rm odd}] \) by
\be\label{eq:H-inp-defn}
	\inp{\tilde Q , Q}_{\ms H} \defn \ms K(\tilde P, P) 
\ee
where $\tilde{P}$ and $P$ are such that \( \tilde Q = \mc K \tilde P \) and \( Q = \mc K P \). That this is indeed an inner product follows from the symmetry, bilinearity and positive definiteness of \(\ms K\) (\autoref{thm:positive-KE}). We can write this inner product in terms of the symplectic form as 
\be
\inp{\tilde Q , Q}_{\ms H} = W[\tilde P, \mc K P] \, .
\label{Hprod2}
\ee

We now complete the space \(\mc K[\ms V_{\rm odd}] \) in the inner product $\inp{\phantom{\_},\phantom{\_} }_{\ms H}$ to obtain a Hilbert space \( \ms H \). Note that $\ms H$ automatically contains all $Q \in \ms V_{\rm even}$ that are of the form $Q = \mc K P$ for $P \in \ms V_{\rm odd}$, and such $Q$ comprise a dense subspace of $\ms H$. In view of \autoref{eq:evol-odd-even}, this means that the even part of all perturbations that are of the form $\Lie_t\delta X$ for some perturbation \(\delta X \in \ms V\) will be represented in $\ms H$. Obviously, unbounded growth of a perturbation $Q$ of the form $ \mc K P$ for $P \in \ms V_{\rm odd}$ suffices to prove instability. However, a perturbation will grow exponentially in time if and only if any of its Lie derivatives with respect to $t^\mu$ grow exponentially in time. As noted in \autoref{rem:2-t-derivative}, the second time derivative of any perturbation \emph{not necessarily in the Lagrangian displacement framework} is in \(\ms V\) and, hence, the even part of its third time derivative is represented in the Hilbert space \(\ms H\). Consequently, stability for $Q$ of the form $ \mc K P$ for $P \in \ms V_{\rm odd}$ implies that no perturbations can grow exponentially --- including those not represented in the Lagrangian framework. However, stability for such $Q$ does not rule out the possibility of instabilities that grow slower than a cubic polynomial in $t$. Thus, stability with respect to perturbations in $\ms H$ is necessary but not sufficient for stability with respect to all perturbations, but it is sufficient to establish mode stability for all perturbations.

	It is convenient to complexify the Hilbert space \(\ms H\) in order to use spectral methods; we will not distinguish this complexification in our notation. The operator $\mc A : \ms H \to \ms H$ given by \autoref{Adef} naturally extends to a real, symmetric operator with dense domain given by the complexification of $\mc K[\ms V_{\rm odd}]$. In particular, it admits a self-adjoint extension $\bar{\mc A}$. 
By a close parallel of the arguments of \cite{PW}, we obtain the following proposition:

\begin{prop}\label{prop:evol}
Given any axisymmetric initial data \(\delta X_0 = (P_0=0, Q_0= \mc K P'_0)\) where $P'_0 \in \ms V_{\rm odd}$ there exists a unique solution $Q(t) \in \mc K[\ms V_{\rm odd}] \subset \ms H$ to \autoref{ddotQ} that is such that \(Q(0) = Q_0\) and \(\dot Q(0) = 0\).
\begin{proof}
Uniqueness of such a $Q(t)$ follows from the spectral arguments given in \cite{PW} (see, in particular, Lemma 6.1 \cite{PW}) while, existence can be shown as follows. Choose a smooth representative ${\tilde P}'_0$ of the \(\ms W_g\)-equivalence class of $P'_0 \in \ms V_{\rm odd}$ and consider the initial data \({\tilde X}'(0) = ({\tilde P}'_0, {\tilde Q}'_0 = 0)\). From \({\tilde X}'(0)\), using \autoref{deltaNs}, we can obtain the initial data \(Y'(0)\) in terms of the physical fluid quantities. From the arguments in \cite{CBYork} and Ch.IX of \cite{CB-book} on the well-posedness of the Einstein-Euler system it follows that, there exists a smooth solution \(Y'(t)\) of the perturbed Einstein-Euler system with initial data \(Y'(0)\). Then using \autoref{eq:dotxi}, we get a smooth solution ${\tilde X}'(t)$ in terms of the Lagrangian displacement with initial data $\tilde{X}'(0)$.

	Now, let $X'(t) = (P'(t), Q'(t))$ denote the \(\ms W_g\)-equivalence class of the solution ${\tilde X}'(t)$ and let $X(t) = (P(t), Q(t))$ denote the \(\ms W_g\)-equivalence class of its time derivative $\Lie_t {\tilde X}'(t)$. It follows from the evolution equations in the form \autoref{eq:evol-op-defn} and \autoref{eq:evol-odd-even} that $Q(t) = \mc K P'(t)$ and that $Q(t)$ satisfies
\be
\frac{d^2Q(t)}{dt^2} = - \mc A Q(t) \, .
\label{QevolH}
\ee
\end{proof}
\end{prop}

By an exact parallel of the proof of Prop.~6.2 of \cite{PW}, we have

\begin{prop} \label{prop:blowup} Let $Q_0 = \mc K P'_0 \in \ms H$ with $P'_0 \in \ms V_{\rm odd}$ be axisymmetric initial data such that the potential energy satisfies $\ms U(Q_0, Q_0) < 0$. Then the solution generated by the initial data $(P_0 = 0, Q_0)$ grows exponentially with time in the sense that there exists $C > 0$ and $\alpha > 0$ such that 
\be
\norm{Q_t}_{\ms H} > C \exp(\alpha t).
\ee
\end{prop}

We also have the following Rayleigh-Ritz-type variational principle to determine the growth rate of the instability. We refer the reader to Theorem 2 \cite{PW} for the proof of this result.
\begin{thm}[Variational Principle for Instability] \label{thm:exp-growth}
For any axisymmetric $P \in {\ms V}_{\rm odd}$ consider the quantity
\be \label{eq:rate-defn}
\omega^2 (P) \defn \frac{{\ms U}({\mc K} P, {\mc K} P)}{{\ms K} (P,P)} \,.
\ee
If $\omega^2 < 0$, the solution $\delta X(t)$ determined by the initial data $(P,0)$ will grow with time at least as fast as $\exp(\alpha t)$ for any $\alpha < |\omega|$, in the sense that the kinetic energy $\ms K$ of $\Lie_t \delta X$ will satisfy 
\be
\lim\limits_{t \to \infty} \left[\ms K (\Lie_t \delta X, \Lie_t \delta X) \exp(-2\alpha t)\right] = \infty \, .
\ee
\end{thm}

\section{Explicit Form of the Variational Principle}\label{sec:algorithm}

In this section, we provide a more concrete form of the variational principle of \autoref{thm:exp-growth}. In the static case, we will provide explicit formulae for the variational principle, and we will show that for spherically symmetric perturbations of a static,
spherically symmetric star, the variational principle reduces to that of  Chandrasekhar \cite{Chandra1, Chandra}, and Seifert and Wald \cite{SW}. In the stationary axisymmetric case, the formulae are too cumbersome to write out explicitly, so we will simply provide an algorithm for performing the calculations needed to evaluate the variational principle.


\subsection{Static Background}

The variational principle \autoref{eq:rate-defn} requires that we compute the potential energy \(\ms U\) corresponding to a \(t\)-reflection-even perturbation of the form \(\mc K P\) where \(P\) is a ``trial function'' consisting of a reflection-odd perturbation $P$. To get an explicit form of the variational principle we first compute an expression for the potential energy for any \(t\)-reflection-even perturbation. In the static case, the reflection-even perturbation takes the form $Q = (0,q_{ab},0,\xi^a)$ (see \autoref{eq:t-decomp}). Using the definition of the canonical energy we have
\be\begin{aligned}
	\mathscr U(Q, Q) & =  \frac{1}{16\pi}\int \df\varepsilon^{(3)}\lb(  p_{ab}\dot q^{ab} - q^{ab} \dot p_{ab} \rb) + \int \left(\dot \xi^\mu {\delta P}_{\mu\nu\lambda\rho} - \xi^\mu \dot{(\delta P)}_{\mu\nu\lambda\rho} - [\xi,\dot\xi]^\mu P_{\mu\nu\lambda\rho}\right).
\end{aligned}\ee
For reflection-even perturbations off of a static background, both $ p_{ab}$ and $\dot \xi^\mu$ vanish (using \autoref{eq:dotxi}), so we obtain
\be\begin{aligned}
	\mathscr U(Q, Q) & =  -\frac{1}{16\pi}\int \df\varepsilon^{(3)}q^{ab} \dot p_{ab} - \int \xi^\mu \dot{(\delta P)}_{\mu\nu\lambda\rho} \\
	& = - \int \df\varepsilon^{(3)} \lb( \frac{1}{16\pi}  q^{ab} \dot p_{ab} + (\rho+p) \xi_a \dot v^a  \rb)
\end{aligned}\ee
where the last equality follows from the same calculation as in \autoref{statxideltaP}. 

Recall that in the case of a static star \(\pi^{ab} = N^a = u^a = 0\). We first compute the gravitational contribution \(-\tfrac{1}{16\pi}\int q^{ab} \dot p_{ab}\) where \(\dot p_{ab}\) is obtained from \autoref{eq:evol-p}. We simplify the resulting expression using the same procedure used in \autoref{sec:positive-KE} for the stationary axisymmetric case to obtain the kinetic energy contribution of the same term. That is, we
\begin{enumerate}
	\item ``Integrate by parts'' any term with two derivatives of \(q_{ab}\) to rewrite it as a quadratic expression in one derivative of \(q_{ab}\).
	\item Rewrite the terms with Ricci tensor, Ricci scalar and two derivatives of the lapse \(N\) using \autoref{eq:constraints-H-ax}, \autoref{eq:Ric-id}, and the linearized Hamiltonian constraint \autoref{hamconstraint} in terms of the background and perturbed matter terms.
\end{enumerate}

Using the above steps we get (omitting the factor of \(\df\varepsilon^{(3)} \) in the intermediate expressions)
\be\begin{split}
	- \int q^{ab}\dot p_{ab} & = \int N \lb[ \tfrac{1}{2} D^c q^{ab} D_c q_{ab} -  D^c q^{ab} D_aq_{bc} - \tfrac{3}{2} D_c q_a{}^a D^c {q_b}^b  + 2 D_a q^{ab} D_b q_c{}^c \rb] \\
	&\quad + 8\pi \int N \lb[ -\tfrac{3}{2} V^{ab} q_{ab} q_c{}^c - 2 V^{ab} q_{ac}q^c{}_b + (E -\tfrac{1}{2} V_c{}^c) (q_a{}^a)^2 - \lb( E - V_c{}^c \rb)q^{ab}q_{ab} \rb] \\
	&\quad + 8\pi \int  N \lb[2\delta E q_a{}^a - \tau^{ab}q_{ab} \rb].
\end{split}\ee
Computing the matter contributions using \autoref{eq:V-fluid-defn} and \autoref{eq:fluid-matter} we get
\be\label{eq:U-grav}\begin{split}
	- \frac{1}{16\pi}\int q^{ab}\dot p_{ab} & = \frac{1}{16\pi} \int N \lb[ \tfrac{1}{2} D^c q^{ab} D^c q_{ab} -  D^c q^{ab} D_aq_{bc} - \tfrac{3}{2} D_c q_a{}^a D^c {q_b}^b  + 2 D_a q^{ab} D_b q_c{}^c \rb] \\
	&\qquad + \int N \lb[ - \tfrac{1}{4}\lb( \rho - p \rb)\lb( q^{ab}q_{ab} + (q_a{}^a)^2\rb) - \tfrac{1}{4}(1-c_s^2)(\rho+p) (q_a{}^a)^2 \rb. \\
	& \qquad\qquad\quad \lb. - \tfrac{1}{2}(2 - c_s^2 )(\rho+p)(D_a\xi^a) q_a{}^a - \tfrac{1}{2} \xi^a D_a(2\rho - p )q_a{}^a \rb] .\\
\end{split}\ee

Next we compute the fluid contribution \(- \int (\rho+p) \xi_a \dot v^a\) using \autoref{eq:dotv} for the static star. Using the Euler equation \autoref{eq:dp-static} we get
\be\begin{split}
	- \int (\rho+p) \xi_a \dot v^a & = \int N \xi_a D^a p \left[ \left(1+c_s^2\right)\lb( \tfrac{1}{2} q_c{}^c + D_c \xi^c \rb) + \frac{1}{\rho+p}\xi^c D_c (\rho+p) \right] \\
	& \qquad - \int N \xi_a D^a \left[c_s^2 (\rho+p) \left(\tfrac{1}{2}q_c{}^c + D_c \xi^c \right) +\xi^c D_c p\right].
\end{split}\ee
Integrating by parts the last term we have
\be\label{eq:U-fluid}\begin{split}
	- \int (\rho+p) \xi_a \dot v^a & = \int N \xi_a D^a p \left[ \lb( \tfrac{1}{2} q_c{}^c + 2D_c \xi^c \rb) + \frac{1}{\rho+p}\xi^c D_c \rho \right] \\
	& + \int N  (D^a \xi_a) \left[c_s^2 (\rho+p) \left(\tfrac{1}{2}q_c{}^c + D_c \xi^c \right) \right] .\\
\end{split}\ee

Putting together \autoref{eq:U-grav} and \autoref{eq:U-fluid} the potential energy for a reflection-even perturbation \(Q\) is
\be\label{eq:U-static}\begin{split}
	\ms U(Q,Q) & = \frac{1}{16\pi} \int \df\varepsilon^{(3)} N \lb[ \tfrac{1}{2} D^c q^{ab} D_c q_{ab} -  D^c q^{ab} D_aq_{bc} - \tfrac{3}{2} D_c q_a{}^a D^c {q_b}^b  + 2 D_a q^{ab} D_b q_c{}^c \rb] \\
	&\quad - \frac{1}{4} \int \df\varepsilon^{(3)} N \lb[ \lb( \rho - p \rb)q^{ab}q_{ab} + \lb( 2p - c_s^2 (\rho+p) \rb)(q_a{}^a)^2 \rb] \\
	& \quad + \int\df\varepsilon^{(3)} N \left[ c_s^2 (\rho+p) (D_a \xi^a)^2  + 2 (D_a \xi^a)(\xi^b D_b p) + \frac{1}{\rho+p}(\xi^a D_a \rho)(\xi^b D_b p) \right] \\
	& \quad - \int \df\varepsilon^{(3)} N \lb[ (1 - c_s^2 )(\rho+p)D_a\xi^a + \xi^a D_a(\rho - p ) \rb]q_b{}^b\,. \\
\end{split}\ee\\

Now we provide an algorithm for computing the variational principle \autoref{eq:rate-defn} in the case of a static star.
\begin{enumerate}[label=(\arabic*), ref=(\arabic*)]
	\item \label{step:P-static} We start with a ``trial" reflection-odd perturbation \(P = (p_{ab},0,v^a,0)\). The linearized Hamiltonian constraint vanishes identically for reflection-odd perturbations, and the linearized momentum constraint \autoref{momconstraint} becomes
\be\label{momconstraintstatic}
D_b p^{ab} = - 8\pi (\rho+p) v^a \, ,
\ee
so we need to start with a solution to this equation.

	One way of generating solutions would be to choose an arbitrary $v^a$ and then solve the elliptic system
\be
D_b D^{(a} Z^{b)} = - 8\pi (\rho+p) v^a \, ,
\ee
for a vector field $Z^a$. One can then choose $p_{ab} = D_{(a} Z_{b)} + {\tilde p}_{ab}$ where ${\tilde p}_{ab}$ is a
solution to $D^b {\tilde p}_{ab} = 0$.
	\item \label{step:KPP-static} We compute \(\ms K(P,P)\) using \autoref{eq:KE-fluid-static}, namely
	\be \label{eq:KPP}
		\ms K (P,P) = \frac{1}{8\pi}\int \df\varepsilon^{(3)} N\lb( p_{ab}p^{ab}-\tfrac{1}{2}({p_a}^a)^2 \rb) + \int \df\varepsilon^{(3)} N (\rho+p) v_a v^a\,.
	\ee
	\item \label{step:Q-KP-static} From the reflection-odd $P$ in \autoref{step:P-static}, we obtain the reflection-even perturbation $Q' = \mathcal K P  = (0, q'_{ab}, 0, \xi'^a)$ using \autoref{Kstatic} i.e.
\be\label{eq:Q-KP}\begin{split}
	q'_{ab} & = 2N\lb( p_{ab}-\tfrac{1}{2}{p_c}^c h_{ab} \rb)\\
	\xi'^a & = N v^a\,.
\end{split}\ee
	\item \label{step:UQQ-static} Compute $\ms U(Q',Q') = \ms U(\mc KP, \mc KP)$ using \autoref{eq:U-static} for the perturbation \(Q'\) from \autoref{step:Q-KP-static} (\autoref{eq:Q-KP}). The variational principle \autoref{eq:rate-defn} takes the form
	\be\label{eq:rate-static}
		\omega^2 = \frac{\ms U(Q',Q')}{\ms K(P,P)}
	\ee
	with the denominator obtained in \autoref{step:KPP-static}.
\end{enumerate}

We now explicitly carry out the above steps in the case of spherically symmetric perturbations of a static, spherically-symmetric star with a ``barotropic'' equation of state of the form $\rho=\rho(n)$. We shall show that our variational principle reduces to that of Chandrasekhar \cite{Chandra1, Chandra}, and Seifert and Wald \cite{SW}. To proceed, following \cite{SW}, it is convenient to work in a gauge where the background metric is
\be
ds^2 = -e^{2 \Psi(r)}dt^2 + e^{2\Lambda(r)}dr^2 +r^2 d\Omega^2
\ee
and where the perturbed spatial metric and the Lagrangian displacement on the Cauchy surfaces \(\Sigma_t\) of constant $t$ are 
\begin{subequations}\label{eq:Q-sph}\begin{align}
	q_{ab} &= 2 e^{2 \Lambda} \lambda(r,t)D_a r D_b r \\
	\xi^a & = \xi(r,t) \left(\frac{\partial}{\partial r}\right)^a
\end{align}\end{subequations}
We can read off that the background lapse is $N=e^{\Psi(r)}$, and we can calculate directly from \autoref{eq:evol-q} and \autoref{eq:dotxi} that 
\begin{subequations}\label{eq:P-sph}\begin{align}
	p_{ab} & = -e^{-\Psi} \dot\lambda r^2 (d\Omega^2)_{ab} \label{eq:p-sph} \\
	v^a & = e^{-\Psi} \dot \xi \left(\frac{\partial}{\partial r}\right)^a \label{eq:v-sph}
\end{align}\end{subequations}
where we have used \((d\Omega^2)_{ab}\) to denote the metric on the unit-radius \(2\)-sphere. The linearized momentum constraint (\autoref{momconstraintstatic}) for \autoref{eq:P-sph} is given by
\be\label{sphereconstraint}
\dot\lambda = -4\pi r e^{2 \Lambda} (\rho+p) \dot\xi \,.
\ee
Unlike in the general case (see \autoref{step:P-static}), for spherically symmetric perturbations the momentum constraint can be solved algebraically. To compare our variational principle with that of \cite{SW} we choose \(\dot\xi\) as a freely specified function on the Cauchy surface \(\Sigma_t\) and use \autoref{sphereconstraint} to substitute for \(\dot\lambda\), and so we have
\begin{subequations}\label{eq:P-sph-alt}\begin{align}
	p_{ab} & = 4\pi e^{2\Lambda-\Psi} (\rho+p) \dot\xi r^3 (d\Omega^2)_{ab}\label{eq:p-sph-alt} \\
	v^a & = e^{-\Psi} \dot \xi \left(\frac{\partial}{\partial r}\right)^a\,. \label{eq:v-sph-alt}
\end{align}\end{subequations}
We use the reflection-odd perturbation \(P=(p_{ab},0,v^a,0)\) from \autoref{eq:P-sph-alt} as our starting ``trial" perturbation to obtain the variational principle; this completes \autoref{step:P-static} of our algorithm. For \autoref{step:KPP-static}, using \autoref{eq:P-sph-alt} in \autoref{eq:KPP} gives
\be\label{eq:K-sph}
\mathscr K(P,P) = 4\pi \int dr\,r^2 e^{3\Lambda-\Psi}(\rho+p)\dot\xi^2.
\ee

The reflection-even perturbation \(Q' = \mc K P\) in \autoref{step:Q-KP-static} can be computed to be
\begin{subequations}\label{eq:Q-KP-sph}\begin{align}
	q'_{ab} & = -8\pi (\rho+p) r e^{4\Lambda} \dot\xi D_a r D_b r \\
	\xi'^a & = \dot \xi \left(\frac{\partial}{\partial r}\right)^a \,.
\end{align}\end{subequations}

	To complete \autoref{step:UQQ-static}, we now substitute the reflection-even perturbation \(Q'\) from \autoref{eq:Q-KP-sph} into \autoref{eq:U-static} and explicitly compute \(\ms U(Q',Q')\). To compare with \cite{SW}, we replace \(\rho\) and \(p\) by the particle number density \(n\) using the identities 
\be
	\rho'(n) = \frac{\rho+p}{n}  \eqsp \rho''(n) = c_s^2 \frac{\rho+p}{n^2}
\ee
which follow from \autoref{pressuredefn} and \autoref{cdefn}, and
\be
	\frac{\partial \rho}{\partial r} = \frac{\rho+p}{n} \frac{\partial n}{\partial r}  \eqsp \frac{\partial p }{\partial r} = c_s^2 \frac{\rho+p}{n}\frac{\partial n}{\partial r}
\ee
which follows from the fact that the background quantities $\rho$, $p$, and $n$ are functions of the single variable $r$.
We simplify the resulting expression by integrating by parts and using the background evolution equations \autoref{eq:Ric-id-static} and \autoref{eq:dp-static} which take the form
\be
\frac{\partial\Psi}{\partial r} + \frac{\partial\Lambda}{\partial r} = 4\pi r e^{2\Lambda}(\rho+p)
\ee
and
\be
c_s^2 \frac{\partial n}{\partial r} = -n \frac{\partial\Psi}{\partial r}.
\ee
We obtain
\be\label{eq:U-sph}\begin{aligned}
\mathscr U(\mathcal KP,\mathcal KP)
= 4\pi \int dr \,  \Biggl[ &\frac{1}{r^2  } \rho''(n) e^{3\Psi+\Lambda}\left(\frac{\partial}{\partial r}\left( r^2 e^{-\Psi} n \dot\xi \right)\right)^2 \\
& - r^2 e^{\Lambda+\Psi} n \rho'(n) \left(\frac{1}{r}+2 \frac{\partial\Psi}{\partial r} \right)\lb(\frac{\partial\Psi}{\partial r} + \frac{\partial\Lambda}{\partial r}\rb) \dot\xi^2  \Biggr] .\\
\end{aligned}\ee

The above expressions \autoref{eq:K-sph} and \autoref{eq:U-sph} for the variational principle agree with those given by Seifert and Wald \cite{SW} modulo the substitution $\dot\xi \mapsto \xi$  (and up to a spurious overall factor of $3$ in both expressions in \cite{SW}). Thus, our variational principle in \autoref{thm:exp-growth} reproduces the Chandrasekhar variational principle \cite{Chandra, Chandra1} for spherical perturbations of a static spherical star with a barotropic equation of state.


\subsection{Stationary Axisymmetric Background}

	Next, we give an algorithm for computing the variational principle \autoref{eq:rate-defn} for axisymmetric perturbations of a stationary-axisymmetric star with circular flow, in parallel with the static case.

\begin{enumerate}[label=(\arabic*), ref=(\arabic*)]
	\item \label{step:P-station} We start with a ``trial" reflection-odd perturbation given by (see \autoref{eq:tphi-decomp})
\be
P = \begin{pmatrix}p_{ab}\\ q_{ab}\\ v^a\\ \xi^a \end{pmatrix} = \begin{pmatrix}\beta_{ab}+\gamma\phi_a\phi_b\\ 2\alpha_{(a}\phi_{b)}\\ v^a\\ 0 \end{pmatrix} 
\ee 
	Note, that we set \(\xi^a = \xi_\parallel\phi^a = 0\) since it is a trivial displacement and does not affect the kinetic energy as discussed in \autoref{sec:positive-KE}. The linearized Hamiltonian constraint vanishes identically for reflection-odd perturbations, and the linearized momentum constraint \autoref{momconstraint} becomes
\be\label{momconstraintstation}
D_b\beta^{ab} -\frac{1}{2}\gamma D^a \Phi +\pi_{b}\lb( 2\Phi D^{[b}\alpha^{a]}-\alpha^b D^a \Phi \rb)  = - 8\pi (\rho+p)\sqrt{1+\Phi U^2} v^a.
\ee
so we need to start with $(\beta_{ab}, \gamma, \alpha_a, v^a)$ (with $\beta_{ab}$, $\alpha_a$ and $v^a$ each orthogonal to $\phi^a$) satisfying this equation.

	Similar to the static case, one way of generating solutions would be to choose an arbitrary $(\gamma, \alpha_a, v^a)$
and then solve the elliptic system
\be
D_b D^{(a} Z^{b)} -\frac{1}{2}\gamma D^a \Phi +\pi_{b}\lb( 2\Phi D^{[b}\alpha^{a]}-\alpha^b D^a \Phi \rb)  = - 8\pi (\rho+p)\sqrt{1+\Phi U^2} v^a
\ee
for an axisymmetric vector field $Z^a$ which is polar i.e. \(Z^a\phi_a = 0\). One can then choose $\beta_{ab} = D_{(a} Z_{b)} + {\tilde \beta}_{ab}$ where ${\tilde \beta}_{ab}$ is a solution to $D^b {\tilde \beta}_{ab} = 0$.
	\item \label{step:KPP-station} We compute \(\ms K(P,P)\) using \autoref{eq:KE-fluid-station} i.e.
	\be\begin{split}
\ms K(P,P) & = \frac{1}{8\pi}\int \df\varepsilon^{(3)}  N\biggl[ \Phi \left(D_{[a}\alpha_{b]}\right)\left(D^{[a}\alpha^{b]}\right) + \beta_{ab}\beta^{ab} + \Phi^2\left(\gamma+2 \alpha_a \pi^a \right)^2 \\
	&\qquad\qquad - \frac{1}{2}\left({\beta_a}^a+\Phi\gamma+2\Phi\alpha_a \pi^a \right)^2 \biggr] + \int \df\varepsilon^{(3)} N(\rho+p) v^a v_a\,.
	\end{split}\ee
	\item \label{step:Q-KP-station} We obtain the reflection-even perturbation
	\be
		Q' = \mathcal K P = \begin{pmatrix}p'_{ab}\\ q'_{ab}\\ v'^a\\ \xi'^a \end{pmatrix} = \begin{pmatrix} 2\Lambda_{(a}\phi_{b)} \\ \Theta_{ab} + \Gamma \phi_{a}\phi_{b} \\ V \phi^a \\ \Xi^a \end{pmatrix}
	\ee
	by using the reflection-odd $P$ of \autoref{step:P-station} in \autoref{eq:K-op-exp}.
	\item \label{step:UQQ-station} Next we need to compute the potential energy  $\ms U(Q',Q') = \ms U(\mc KP, \mc KP)$ for the perturbation \(Q'\) from \autoref{step:Q-KP-station}. The explicit formula for the potential energy in this case is cumbersome to write out but can be computed as follows.
	\be\begin{split}\label{Ustationary}
	\ms U(Q',Q') & = \frac{1}{16\pi}\int \df\varepsilon^{(3)} \lb[ p'_{ab} \dot{(q'^{ab})} - q'^{ab} \dot{(p'_{ab})} \rb] \\
	&\quad + \int \left[ \dot{(\xi'^\mu)} \delta' P_{\mu\nu\lambda\rho}  -  \xi'^\mu \dot{({\delta' P})}_{\mu\nu\lambda\rho} - [\xi',\dot{(\xi')}]^\mu P_{\mu\nu\lambda\rho} \right]
	\end{split}\ee
where the quantities $p'_{ab}$, $q'_{ab}$, $v'_a$, and $\xi'^a$ are obtained from \autoref{step:Q-KP-station}, and by substituting these into the evolution equations \autoref{eq:evol-pq} and \autoref{eq:evol-xiv} one can obtain expressions for $\dot{(p'_{ab})}$, $\dot{(q'_{ab})}$, $\dot {(v'_a)}$, and $\dot{(\xi'^a)}$. In evaluating $\dot{(p'_{ab})}$, one will need to calculate the perturbed energy density and pressure under the perturbation $Q'$ in order to calculate $\tau'^{ab}=\delta'T^{ab}$. From \autoref{deltarhoandp}, these are
\be\begin{aligned}
\delta'\rho &= (\rho+p) \frac{\Delta' n}{n}-\xi'^a D_a \rho \\
\delta' p &= c_s^2(\rho+p) \frac{\Delta' n}{n}-\xi'^a D_a p,\\
\end{aligned}\ee
where $(\Delta' n)/n$ is given by \autoref{Deltanovern} (substituting $Q'$ in). Similarly, these are needed to evaluate the quantity $\delta' P_{\mu\nu\lambda\rho}$ appearing in the third term of \autoref{Ustationary} from the definition \autoref{eq:P-defn} of $P_{\mu\nu\lambda\rho}$. 
By a calculation similar to \autoref{stationaryPcalc}, the fourth term of \autoref{Ustationary} is
\be\label{eq:fourth-term}
-  \int \xi'^\mu \dot{({\delta' P})}_{\mu\nu\lambda\rho} = -\int \sqrt{1+\Phi U^2}(\rho+p)\xi'^a \left[ \dot {(v'_a)} + U \phi^b \dot {(q'_{ab})} \right], 
\ee
while the final term is
\be\label{eq:final-term}
-\int\left[ \xi', \dot{(\xi')} \right]^\mu P_{\mu\nu\lambda\rho}  = \int  \sqrt{1+\Phi U^2}(\rho+p) \Phi U  \xi'^a D_a \left(\Phi^{-1} \phi_b \dot{(\xi'^b)}\right)
\ee
where we have used the fact that $\dot{(\xi'^a)}$ is axial (i.e., tangent to $\phi^a$) and the axisymmetry of $\xi'^a$. In \autoref{eq:fourth-term}-\autoref{eq:final-term}, we need to substitute for $\dot{(q'_{ab})}$, $\dot {(v'_a)}$, and $\dot{(\xi'^a)}$ using \autoref{eq:evol-pq} and \autoref{eq:evol-xiv}.

Finally, the variational principle \autoref{eq:rate-defn} takes the form
	\be\label{eq:rate-station}
		\omega^2 = \frac{\ms U(Q',Q')}{\ms K(P,P)}
	\ee
	with the denominator obtained in \autoref{step:KPP-station}.
\end{enumerate}

Using the metric ansatz of \cite{CF2}, a tedious computation shows that the above algorithm reproduces the variational formula of \cite{CF2} for the frequency of modes. Note, however that \cite{CF2} did not show the positivity of the kinetic energy term and thus did not obtain a Rayleigh-Ritz variational principle.

\section*{Acknowledgements}

We thank John Friedman for helpful comments on the initial draft of the paper. This research was supported in part by the NSF grants PHY~12-02718 and PHY~15-05124 to the University of Chicago. JSS is supported by ERC grant ERC-2011-StG 279363-HiDGR.

\appendix

\section{Fluid Evolution Equations}\label{fluideqs}

In this appendix we compute the evolution equations for the perturbed fluid initial data variables, $(v^a, \xi^a)$, of a perturbation off of a stationary background. We assume that $\xi^\mu$ and its time derivatives have been made tangent to $\Sigma$ by the addition of an appropriate flowline trivial. As in the linearized ADM equations, \autoref{eq:evol-pq}, there will be terms coming from gauge transformations generated by the perturbed lapse and shift. We ignore these in the intermediate computations by setting the perturbed lapse and shift to zero and insert them back in the final expressions. With these choices we have, on $\Sigma$,
\be
\delta g_{\mu\nu} = {h^a}_\mu {h^b}_\nu q_{ab} \eqsp v^a = \delta u^a = \delta \left( {h^a}_\mu u^\mu\right) = {h^a}_\mu \delta u^\mu.
\ee
Using \autoref{deltau} we then find
\be
 v^a = \frac{1}{2} u^a u^b u^c q_{bc}+  \left[({\delta^a}_b + u^a u_b){h^b}_\mu+\sqrt{1+u^2}\, u^a  \nu_\mu \right] \Lie_u \xi^\mu.
\ee
Using the fact that we can write
\be
u^\mu = \frac{\sqrt{1+u^2}}{N}(t^\mu - N^\mu) + {h^\mu}_a u^a
\ee 
as well as
\be
\nu_\mu \Lie_u \xi^\mu = - \xi^\mu \Lie_u \nu_\mu =  \xi^\mu \Lie_u (N \nabla_\mu t) = N\xi^\mu \nabla_\mu \Lie_u   t = N\xi^a D_a \left(\frac{\sqrt{1+u^2}}{N} \right)
\ee
we find
\be\begin{aligned}
 v^a &= \frac{1}{2} u^a u^b u^c q_{bc} +  \left({\delta^a}_b + u^a u_b\right)\left(\frac{\sqrt{1+u^2}}{N}\left[   \dot\xi^b - \Lie_N \xi^b \right] + \Lie_u \xi^b\right)\\
 &\qquad+\sqrt{1+u^2}\, u^a  N\xi^b D_b \left(\frac{\sqrt{1+u^2}}{N} \right).\\
\end{aligned}\ee
Inverting this to solve for $\dot\xi^a$ gives the first evolution equation:
\be
\begin{aligned} \label{dotxi1}
\dot\xi^a =& \Lie_N \xi^a - \tfrac{N}{\sqrt{1+u^2}} \Lie_u \xi^a\\
  &+\tfrac{N}{\sqrt{1+u^2}}\left({\delta^a}_d -\tfrac{1}{1+u^2} u^a u_d\right) \left[v^d - \tfrac{1}{2} u^d u^b u^c q_{bc} -N\sqrt{1+u^2}\, u^d  \xi^b D_b \left(\tfrac{\sqrt{1+u^2}}{N} \right) \right].\\
\end{aligned}
\ee
We add back a gauge transformation $\hat\xi^a$ generated by the perturbed lapse and shift to get \autoref{eq:dotxi}.

To get the evolution equation for $v^a$, we linearize the Euler equation \autoref{eq:euler-eqn} off of a stationary background, obtaining
\be\begin{aligned}\label{vdot1}
\dot v^a =&  \Lie_N v^a - \frac{\left(q_{bc}u^b u^c + 2v^b u_b\right)D^a N}{2\sqrt{1+u^2}}  + \sqrt{1+u^2}\,q^{ab} D_b N  +N\left(u^a{p^b}_b  -2 u_b p^{ab}\right)\\
&  +\tfrac{N}{\sqrt{h}}\left(  \pi^{ab} u_b{q^c}_c -2\pi^{ab}u^c q_{bc}
 -\tfrac{1}{2} {\pi^b}_b u^a{q^c}_c   + \pi^{bc} u^a q_{bc}+{\pi^b}_b v^a -2 \pi^{ab}v_b\right) 
 \\
&+\frac{N(q_{cd}u^c u^d + 2v^c u_c)u^b D_b u^a}{2(1+u^2)^{3/2}} - \frac{N(v^b D_b u^a+ u^b D_b v^a)}{\sqrt{1+u^2}} \\
& - \frac{Nu^b u^c  \left(2 D_b {q_c}^a - D^a q_{bc}  \right)}{2\sqrt{1+u^2}}  + \left(  \frac{N\left(D^a p+ u^a u^bD_b p\right)}{\sqrt{1+u^2}} -u^a N^b D_b p  \right)\frac{\delta\rho+\delta p}{(\rho+p)^2}
\\
& +\frac{N(q_{bc}u^b u^c + 2v^b u_b)\left(D^a p+ u^a u^b D_b p\right)}{2(\rho+p)(1+u^2)^{3/2}} + \frac{v^a  N^b D_b p -u^a  \dot{(\delta p)} +u^a N^bD_b \delta p}{\rho+p} 
\\
&  -\frac{N\left(D^a \delta p- q^{ab} D_b p+ 2 v^{(a} u^{b)}D_b p+u^a u^b D_b \delta p\right)}{(\rho+p)\sqrt{1+u^2}}  \,.\\
\end{aligned}\ee
This is not the desired evolution equation, since $\dot{(\delta p)}$ --- which depends on $\dot v^a$ --- appears on the right-hand-side. From \autoref{deltarhoandp}, we have
\be\label{dotdeltap}
\dot{(\delta p)} = c_s^2 (\rho+p) \frac{\dot{(\Delta n)}}{n} - \dot\xi^a D_a p.
\ee
To evaluate $\dot{(\Delta n)}/n$, first note that $\sqrt{h} n \sqrt{1+u^2}$ is equal (up to a factor of a fixed non-dynamical volume element) to the pullback of $\boldsymbol N$ to $\Sigma$, and therefore has vanishing Lagrangian perturbation, giving
\be\begin{aligned} \label{Deltan}
\frac{\Delta n}{n} &= -\frac{\Delta\sqrt{h}}{\sqrt{h}} - \frac{\Delta\sqrt{1+u^2}}{\sqrt{1+u^2}}
\\
&= -\frac{1}{2}({q_a}^{a} + 2 D_a \xi^a) - \frac{1}{\sqrt{1+u^2}} \left( \frac{u^a u^b q_{ab} + 2 u_a v^a}{2\sqrt{1+u^2}} + \xi^a D_a \sqrt{1+u^2} \right)
\\
&=
-\frac{1}{2}\left(h^{ab} + \frac{u^a u^b}{1+u^2}\right)q_{ab} - \frac{ u_a v^a}{1+u^2}-\frac{D_a\left(\sqrt{1+u^2}\, \xi^a\right)}{\sqrt{1+u^2}} 
\end{aligned}\ee
so that
\be
\frac{\dot{(\Delta n)}}{n} =  -\frac{1}{2}\left(h^{ab} + \frac{u^a u^b}{1+u^2}\right)\dot q_{ab} - \frac{ u_a \dot v^a}{1+u^2}-\frac{D_a\left(\sqrt{1+u^2}\, \dot\xi^a\right)}{\sqrt{1+u^2}} .
\ee

Substituting this into \autoref{dotdeltap}, then into \autoref{vdot1}, and solving for $\dot v^a$ gives
\be\begin{aligned}
 \dot v^a & = \left({\delta^a}_e + \tfrac{c_s^2}{1+(1-c_s^2)u^2} u^a u_e\right) \times \Biggl\{
   \Lie_N v^e - \tfrac{1}{2\sqrt{1+u^2}}\left(q_{bc}u^b u^c + 2v^b u_b\right)D^e N  \\
	&\quad + \sqrt{1+u^2}\,q^{eb} D_b N  +N\left(u^e{p^b}_b  -2 u_b p^{eb}\right)- \tfrac{N}{2\sqrt{1+u^2}}u^b u^c  \left(2 D_b {q_c}^e - D^e q_{bc}  \right) \\
	&\quad  +\tfrac{N}{\sqrt{h}}\left(  \pi^{eb} u_b{q^c}_c -2\pi^{eb}u^c q_{bc}
 -\tfrac{1}{2} {\pi^b}_b u^e{q^c}_c   + \pi^{bc} u^e q_{bc}+{\pi^b}_b v^e -2 \pi^{eb}v_b\right) \\
	&\quad +\tfrac{N}{2(1+u^2)^{3/2}}(q_{cd}u^c u^d + 2v^c u_c)u^b D_b u^e - \tfrac{N}{\sqrt{1+u^2}}(v^b D_b u^e+ u^b D_b v^e) \\
	&\quad - \frac{1+c_s^2}{\rho+p}  \left(-\frac{\Delta n}{n}  + \frac{\xi^c D_c \ln(\rho+p)}{1+c_s^2}\right) \left( \tfrac{N}{\sqrt{1+u^2}}\left(h^{eb}+ u^e u^b \right) -u^e N^b \right) D_bp \\
	&\quad  +\frac{N(q_{bc}u^b u^c + 2v^b u_b)\left(D^e p+ u^e u^b D_b p\right)}{2(\rho+p)(1+u^2)^{3/2}} + \frac{v^e  N^b D_b p}{\rho+p}  +\frac{N\left( q^{eb} D_b p- 2 v^{(e} u^{b)}D_b p\right)}{(\rho+p)\sqrt{1+u^2}}  \\ 
	&\quad - \left(\frac{\sqrt{1+u^2}\,u^e N^b-N\left(h^{eb} +u^e u^b \right)}{(\rho+p)\sqrt{1+u^2}}\right) D_b \left[- c_s^2 (\rho+p) \frac{\Delta n}{n} +\xi^c D_c p\right]\\
	&\quad   +\frac{1}{2}c_s^2 u^e\left(h^{bc} + \frac{u^b u^c}{1+u^2}\right)\dot q_{bc} + \frac{c_s^2 u^e D_b\left(\sqrt{1+u^2}\, \dot\xi^b\right)}{\sqrt{1+u^2}}  + \frac{u^e \dot \xi^b D_b p}{(\rho+p)}   \Biggr\}.
\end{aligned}\ee
where we have used \autoref{deltarhoandp} and \(\Delta n / n\) is given by \autoref{Deltan}. To obtain an explicit formula in terms of the initial data $(p_{ab}, q_{ab}, v^a, \xi^a)$, it remains to substitute \autoref{eq:evol-q} for $\dot q_{ab}$ and \autoref{eq:dotxi} for $\dot \xi^a$ into the final line. Note that when the fluid star is not static, it is necessary to have $c_s^2 \leq 1$ to solve for $\dot v^a$, otherwise the factor $c_s^2/[1+(1-c_s^2)u^2]$ can diverge. Adding a gauge transformation $\hat v^a$, generated by the perturbed lapse and shift then gives \autoref{eq:dotv}.

\newpage

\bibliographystyle{JHEP}
\bibliography{KE}       
\end{document}